\documentclass[preprint,10pt,3p]{elsarticle}

\usepackage{graphics}
\usepackage{epsfig}
\usepackage{times}
\usepackage{amsmath}
\usepackage{amssymb}
\usepackage{epstopdf}
\usepackage{subfigure,color,balance}
\usepackage{caption}
\usepackage{verbatim}
\usepackage{cases}
\allowdisplaybreaks[4]
\usepackage{enumerate}
\usepackage{booktabs}
\usepackage{balance}
\usepackage{mathbbol}
\usepackage{multirow}
\usepackage[ruled,linesnumbered]{algorithm2e}
\usepackage{algpseudocode}
\usepackage{mathrsfs}
\usepackage{threeparttable}
\usepackage[draft]{todonotes}
\usepackage{makecell}
\usepackage{siunitx}
\usepackage{diagbox}
\usepackage{threeparttable}
\usepackage{colortbl}
\usepackage{enumitem}

\usepackage[colorlinks=true,      
linkcolor=black,      
citecolor=black,      
filecolor=black,      
urlcolor=blue]{hyperref}

\newtheorem{definition}{Definition}

\newdefinition{remark}{Remark}
\newtheorem{lemma}{Lemma}

\newtheorem{assumption}{Assumption}

\def\g{{\bf g}}

\def\x{{\bf x}}

\def\0{{\bf 0}}
\def\1{{\bf 1}}

 \biboptions{compress}

\begin{document}
	
	\begin{frontmatter}
		
		\title{{Robust Nonlinear Data-Driven Predictive Control for Mixed Vehicle Platoons via Koopman Operator and Reachability Analysis}} 

        \author[a]{Shuai Li}
		\ead{li-s21@mails.tsinghua.edu.cn}
	\author[b]{Jiawei Wang}
		\ead{jiawe@umich.edu}
        \author[c]{Kaidi Yang\corref{cor1}}
		\ead{ykaidi@nus.edu.sg}
        \author[a]{Qing Xu}
		\ead{qingxu@tsinghua.edu.cn}
        \author[a]{Jianqiang Wang}
		\ead{wjqlws@tsinghua.edu.cn}
        \author[a]{Keqiang Li\corref{cor1}}
		\ead{likq@tsinghua.edu.cn}
		
		\cortext[cor1]{Corresponding author: Kaidi Yang and Keqiang Li}

		\address[a]{School of Vehicle and Mobility, Tsinghua University, Beijing 100084, China}
		\address[b]{Department of Civil and Environmental Engineering, University of Michigan, Ann Arbor, MI 48109, USA}
            \address[c]{Department of Civil and Environmental Engineering, National University of Singapore, Singapore 117576, Singapore}
		
\begin{abstract}
Mixed vehicle platoons, comprising connected and automated vehicles (CAVs) and human-driven vehicles (HDVs), hold significant potential for enhancing traffic performance. However, most existing control strategies assume linear system dynamics and often ignore the impact of adverse factors such as noise, disturbances, and attacks, which are inherent to real-world scenarios. To address these limitations, we propose a Robust Nonlinear Data-Driven Predictive Control (RNDDPC) framework that ensures safe and optimal control under uncertain and adverse conditions. By utilizing Koopman operator theory, we map the system's nonlinear dynamics into a higher-dimensional space, constructing a Koopman-based model that approximates the behavior of the original nonlinear system. To mitigate modeling errors associated with this predictor, we introduce a data-driven reachable set analysis technique that performs secondary learning using matrix zonotope sets, generating a reachable set predictor for over-approximation of the future states of the underlying system. Then, we formulate the RNDDPC optimization problem and solve it in a receding horizon manner for robust control inputs. Extensive simulations demonstrate that the proposed framework significantly outperforms baseline methods in tracking performance under noise, disturbances, and attacks. 
\end{abstract}
		
		
	\end{frontmatter}

\section{Introduction}

\subsection{Background and Literature Review}
With the rapid advancement of information and communication technologies, connected and automated vehicles (CAVs) have achieved significant progress in mass production and deployment, steadily gaining a growing share of the automotive market~\cite{hu2025safety,wang2023new}. As this trend continues, traffic systems are poised to transition into mixed traffic environments, where CAVs and human-driven vehicles (HDVs) coexist~\cite{chen2021mixed,shi2021connected,li2023survey}. While these developments hold great potential for transforming transportation systems, ensuring driving safety and improving traffic efficiency in mixed traffic scenarios remain critical challenges~\cite{amoozadeh2015platoon,zhu2020synthesis}. Mixed vehicle platoons, comprising both CAVs and HDVs, have garnered increasing attention for their potential to address these challenges. The feasibility and effectiveness of such platoons have been validated through traffic simulations~\cite{li2025influence,wang2023deep} and real-world experiments~\cite{jin2018experimental,hajdu2019robust}.

Research on mixed vehicle platoon control strategies has primarily focused on model-based approaches, which rely on well-established car-following models to describe the behavior of HDVs. Commonly used models include the optimal velocity model (OVM)~\cite{bando1995dynamical} and the intelligent driver model (IDM)~\cite{treiber2000congested}. These models provide the foundation for state-space representations of mixed vehicle platoons, enabling the design of various control strategies. Control techniques such as $\mathcal{H}_\infty$ robust control~\cite{wang2022robustness}, linear quadratic regulator~\cite{jin2018experimental}, adaptive control~\cite{tang2024adaptive}, model predictive control (MPC)~\cite{feng2021robust}, and control barrier function~\cite{zhao2023safety} have been proposed to achieve optimal or near-optimal performance with theoretical guarantees. However, the performance of model-based strategies heavily depends on the accuracy of the underlying models. Accurately identifying parameters in these models, particularly for HDVs, poses a significant challenge and may limit the practical effectiveness of such control approaches.

In response to the limitations of model-based approaches, data-driven methods have gained increasing attention for their ability to bypass the explicit parameter identification required in system models. Among these methods, data-driven MPC integrates the established MPC framework with data-driven techniques, showing potential for achieving optimal control while respecting system constraints and maintaining stability guarantees~\cite{berberich2020data,wang2024learning}. One prominent example is the application of data-enabled predictive control (DeePC)~\cite{coulson2019data} to mixed vehicle platoons. DeePC and its variants have been validated through simulation studies, showing their effectiveness in dampening traffic waves~\cite{wang2023deep}, reducing energy consumption~\cite{li2024physics}, and protecting privacy safety~\cite{zhang2023privacy}. However, deploying these strategies in real-world traffic scenarios faces several challenges. Noise inherent in onboard sensors and roadside perception systems can compromise data accuracy~\cite{lan2021data}. The dynamic and often unpredictable variations of the head vehicle's velocity create mismatches between online predictions and actual traffic conditions~\cite{shang2023decentralized}. Additionally, the adoption of Internet of Vehicles (IoV) technologies exposes CAV control systems to potential adversarial attacks, such as maliciously altered control inputs or manipulated sensor data~\cite{xu2022reachability}, which could compromise safety. Despite these concerns, most existing studies fail to adequately account for the influence of noise in data collection and online control, oversimplistically assume constant head vehicle velocities, and neglect the risks posed by attacks. These simplifications can reduce trajectory tracking accuracy, weaken robustness, and increase safety risks in mixed vehicle platoon control.

Recent studies have made significant progress in enhancing the robustness of mixed vehicle platoon systems. For instance, reformulating the DeePC framework using min-max robust optimization has been shown to mitigate the impact of unknown disturbances originating from the head vehicle~\cite{shang2023decentralized}. A distributed data-driven MPC approach with feedforward compensation has also effectively countered external disturbances~\cite{guo2022distributed}. Additionally, zonotopic data-driven predictive control (ZPC), which utilizes linear reachability analysis, has successfully minimized the influence of observational noise on data collection~\cite{lan2021data}. While these approaches enhance robustness, they exhibit certain limitations. Under noise-free conditions, the DeePC predictor established in~\cite{shang2023decentralized} has been proven to be equivalent to the linear MPC formulation for linear time-invariant (LTI) systems~\cite{coulson2019data,lazar2024basis}. Similarly, the subspace identification matrix generation in~\cite{guo2022distributed} depends on the assumption of linearity, and the reachability set computation in~\cite{lan2021data} also relies on the properties of linear systems. In practice, mixed vehicle platoons exhibit strongly nonlinear behavior, particularly in human driver responses. The aforementioned methods do not fully capture these nonlinear dynamics, limiting their effectiveness in real-world applications.

\subsection{Contributions}

To address these gaps concerning robustness and nonlinearity in mixed vehicle platoons, we propose a Robust Nonlinear Data-Driven Predictive Control (RNDDPC) framework that integrates Koopman operator theory and reachability analysis. Specifically, considering the nonlinear nature of mixed vehicle platoons, inspired by~\cite{zhan2022data,korda2018linear,korda2020koopman}, we employ Koopman operator theory to represent the nonlinear system as a high-dimensional linear system. Based on the extended dynamic mode decomposition (EDMD) method~\cite{williams2015data}, we effectively utilize a deep neural network (DNN) to learn and construct a Koopman-based model that approximates the dynamics of the original nonlinear system. Notably, unlike previous methods~\cite{zhan2022data,korda2018linear,korda2020koopman} directly apply the Koopman-based model for control, the proposed method explicitly addresses the challenges arising from adverse factors, including noise, disturbances, and attacks in real-world applications. To enhance robustness, we introduce a data-driven reachability analysis method, utilizing matrix zonotope sets to perform secondary learning and generate a Koopman-based reachable set predictor. This predictor over-approximates the system’s future states, accounting for uncertainties and ensuring robust predictions under various conditions. Within a receding horizon manner, we formulate and solve the RNDDPC optimization problem with safety constraints to compute robust control inputs. This enables the development of a robust data-driven predictive control strategy for nonlinear mixed vehicle platoons, ensuring system safety against uncertain and adverse conditions. Precisely, the key contributions of this work are as follows:

\begin{itemize}

\item  We introduce a Koopman-based deep EDMD method to address the system nonlinearity in the modeling of mixed vehicle platoons. Unlike previous work~\cite{lan2021data,shang2023decentralized,guo2022distributed} that relies on linear assumptions, our approach explicitly captures the nonlinear characteristics of the system. By utilizing actual measurement data, the proposed method approximates the lifting function via DNN, generating a high-dimensional Koopman-based model that effectively represents the dynamics of mixed vehicle platoons. This eliminates the need for explicit model knowledge, which is often required by methods~\cite{jin2018experimental,wang2022robustness,feng2021robust,zhao2023safety}. The linearity of the Koopman-based model facilitates controller design using well-established linear system theory, offering a more efficient and interpretable solution compared to complex nonlinear modeling strategies or less transparent reinforcement learning techniques.

\item We propose the RNDDPC method that integrates a robust control method with the Koopman-based reachable set predictor to explicitly handle adverse factors, including modeling errors, disturbances, and attacks. Unlike previous approaches~\cite{zhan2022data,korda2018linear,korda2020koopman} that directly apply the Koopman-based model for control design, our method addresses modeling errors caused by noise and inaccuracies. To enhance robustness, we utilize matrix zonotope set technology for secondary learning, constructing an over-approximated set of system models. Additionally, we account for potential adverse factors by defining them as zonotope sets, rather than assuming them to be zero as in~\cite{wang2023deep}. For each control horizon, we formulate a Koopman-based reachable set predictor using matrix zonotope set technology and design an RNDDPC optimization problem to ensure that the predicted states always satisfy safety constraints, achieving safe and optimal control for CAVs.

\item We extensively evaluate the effectiveness of the proposed RNDDPC method through extensive simulations, with a particular focus on its robustness against adverse factors, including noise, disturbances, and attacks. A comparative analysis is conducted with multiple baseline approaches, including linear MPC, nonlinear MPC, Koopman MPC~\cite{korda2018linear,korda2020koopman}, DeePC~\cite{wang2023deep,coulson2019data}, and ZPC~\cite{lan2021data,alanwar2022robust}. The simulation results indicate that, compared to all baselines, RNDDPC significantly reduces the tracking velocity mean error, tracking spacing mean error, and total real cost in both comprehensive and emergency scenarios. These results highlight the enhanced robustness and tracking accuracy of the RNDDPC method for controlling mixed vehicle platoons, particularly in the presence of adverse conditions.

\end{itemize}

\subsection{Paper Organization}
The rest of this paper is organized as follows. Section~\ref{Sec:2} provides the foundational preliminaries. Section~\ref{Sec:3} presents data-driven modeling of mixed vehicle platoon systems. Section~\ref{Sec:4} introduces the formulation of the proposed RNDDPC framework. Numerical simulations are provided in Section~\ref{Sec:5}, and Section~\ref{Sec:6} concludes this paper.

\section{Preliminaries}
\label{Sec:2}
This section presents the essential preliminaries on Koopman operator and reachable sets for data-driven predictive control. For clarity and conciseness, we may slightly abuse some notations, which will be defined specifically for use within this section.

\subsection{Basics of Koopman Operator}
We begin by briefly introducing the Koopman operator framework, which provides a linear representation of nonlinear dynamical systems in a high-dimensional function space. This property allows the inherently nonlinear dynamics of the mixed vehicle platoon system, often difficult to model and control using conventional methods, to be transformed into a linear form that is more amenable to analysis and control design.

\begin{definition}[Koopman Operator~\cite{koopman1931hamiltonian}]
	\label{Definition:KoopmanOperator}
	Consider a discrete-time, autonomous, nonlinear dynamical system described by
        \begin{equation}
        \label{BasicEq:KoopmanOperator1}
        {x}(k+1)=f\left(x\left(k\right)\right),
        \end{equation}	 
        where ${x}(k) \in \mathbb{R} ^{n}$ denotes the system state at time step $k$, and $f(\cdot)$ is a nonlinear state transition function. The Koopman operator $\mathcal{K} \in \mathbb{R} ^{h \times h}$ is defined as a linear operator that acts on a space of observable functions $h(x(k)) \in \mathbb{R} ^{h}$. It governs the evolution of these observables according to
        \begin{equation}
        \label{BasicEq:KoopmanOperator2}
        h(x(k+1))=\mathcal{K}h(x(k)),
        \end{equation}
        enabling the original nonlinear system~\eqref{BasicEq:KoopmanOperator1} to be equivalently represented in a lifted linear form~\eqref{BasicEq:KoopmanOperator2}.
\end{definition}

The Koopman operator framework effectively captures the essential characteristics of nonlinear systems by transforming them into a higher-dimensional space, where their dynamics can be described linearly. This property makes it particularly attractive for data-driven modeling and control design. For further details on Koopman operator theory and its applications, readers may refer to~\cite{korda2018linear,korda2020koopman}.

\subsection{Basics of Reachable Sets}
Reachability analysis is a powerful tool for ensuring safety in control systems, however, its practical application is often limited by high computational complexity. In mixed vehicle platoon control, where real-time performance is critical, efficient computation is essential. To address this, and inspired by~\cite{lan2021data,xu2022reachability,alanwar2022robust}, we employ zonotope sets to represent reachable sets, significantly improving computational efficiency. The key definitions and set representations relevant to reachable set computations are outlined below.
\begin{definition}[Interval Set~\cite{althoff2010reachability}]
	\label{Definition:IntervalSet}
	An interval set $ \mathcal{H}$ is a connected subset of $\mathbb{R} ^{n}$, and it can be defined as $ \mathcal{H}=\left\{x_\mathcal{H} \in \mathbb{R}^{n} \mid \underline{x}_{\mathcal{H}_i} \leq x_{\mathcal{H}_i} \leq \overline{x}_{\mathcal{H}_i}, \forall i=1, \ldots, n\right\}$, where $ \underline{x}_{\mathcal{H}_i} $ and $\overline{x}_{\mathcal{H}_i}$ are the lower bound and upper bound of $x_{\mathcal{H}_i}$, respectively. Interval set can be defined as $\mathcal{H}=[\underline{\mathcal{H}},\overline{\mathcal{H}}]$, with $\underline{\mathcal{H}} =[\underline{x}_{\mathcal{H}_1},\underline{x}_{\mathcal{H}_2},\ldots,\underline{x}_{\mathcal{H}_n}]$
	and $\overline{\mathcal{H}}= [\overline{x}_{\mathcal{H}_1},\overline{x}_{\mathcal{H}_2},\ldots,\overline{x}_{\mathcal{H}_n}]$.
\end{definition}

\begin{definition}[Zonotope Set~\cite{kuhn1998rigorously}]
	\label{Definition:ZonotopeSet}
	Given a center vector $c_\mathcal{Z} \in \mathbb{R} ^n$, and $\gamma_\mathcal{Z} \in \mathbb{N}$ generator vectors in a generator matrix $G_\mathcal{Z} =\left[g_\mathcal{Z}^{(1)}, g_\mathcal{Z}^{(2)},\ldots, g_\mathcal{Z}^{(\gamma_\mathcal{Z})}\right] \in \mathbb{R}^{n \times \gamma_\mathcal{Z}}$, a zonotope set is defined as 
    \begin{equation}
		\label{Eq:Zonotope}
		\mathcal{Z} =\left \langle c_\mathcal{Z}, G_\mathcal{Z}\right \rangle = \left\{x \in \mathbb{R}^n \mid x=c_\mathcal{Z}+\sum_{i=1}^{\gamma_\mathcal{Z}} \beta^{(i)} g_\mathcal{Z}^{(i)},-1 \leq \beta^{(i)} \leq 1\right\}.
    \end{equation}
  For zonotope sets, the following operations hold:
	\begin{itemize}
		\item  \textit{Linear Map:} For a zonotope set $\mathcal{Z}=\left \langle c_\mathcal{Z}, G_\mathcal{Z}\right \rangle $, $L \in \mathbb{R}^{m\times n}$, the linear map is defined as $L\mathcal{Z}=\left \langle Lc_\mathcal{Z}, LG_\mathcal{Z}\right \rangle$.
		
		\item  \textit{Minkowski Sum:} Given two zonotope sets $\mathcal{Z}_1=\left \langle c_{\mathcal{Z}_1}, G_{\mathcal{Z}_1}\right \rangle$ and $\mathcal{Z}_2=\left \langle c_{\mathcal{Z}_2}, G_{\mathcal{Z}_2}\right \rangle$ with compatible dimensions, the Minkowski sum is defined as $\mathcal{Z}_1 + \mathcal{Z}_2=\left\langle c_{\mathcal{Z}_1}+c_{\mathcal{Z}_2},\left[G_{\mathcal{Z}_1}, G_{\mathcal{Z}_2}\right]\right\rangle$. 
		
		\item  \textit{Cartesian Product:} Given two zonotope sets $\mathcal{Z}_1=\left \langle c_{\mathcal{Z}_1}, G_{\mathcal{Z}_1}\right \rangle$ and $\mathcal{Z}_2=\left \langle c_{\mathcal{Z}_2}, G_{\mathcal{Z}_2}\right \rangle$, the cartesian product is defined as
		\vspace{-0.1cm}
		\begin{equation}
		\label{Eq:Cartesian}
		\mathcal{Z}_1 \times\mathcal{Z}_2  =\left\langle\begin{bmatrix}
		c_{\mathcal{Z}_1} \\
		c_{\mathcal{Z}_2}
		\end{bmatrix},\begin{bmatrix}
		G_{\mathcal{Z}_1} & 0 \\
		0 & G_{\mathcal{Z}_2}
		\end{bmatrix}\right\rangle.
		\end{equation}
		
		\item  \textit{Over-approximation:} A zonotope set 	$\mathcal{Z}=\left \langle c_\mathcal{Z}, G_\mathcal{Z}\right \rangle$ could be over-approximated by a interval set denoted as Definition~\ref{Definition:IntervalSet} with operation is defined as
		\begin{equation}
		\label{Eq:over-approximation}
		\mathcal{H} = \mathsf{interval}(\mathcal{Z}) = [c_\mathcal{Z}-\bigtriangleup {g_\mathcal{Z}},c_\mathcal{Z}+\bigtriangleup {g_\mathcal{Z}}],
		\end{equation}
		where $\bigtriangleup {g_\mathcal{Z}} = \sum_{i=1}^{\gamma_\mathcal{Z}}\left|g_\mathcal{Z}^{(i)}\right|$, with the absolute value is taken element-wise. 
	\end{itemize}
\end{definition}

\begin{definition}[Matrix Zonotope Set~\cite{althoff2010reachability}]
\label{Definition:MatrixZonotopeSet}
	Given a center matrix $C_\mathcal{M} \in \mathbb{R} ^{n \times m}$, and $\gamma_\mathcal{M} \in \mathbb{N}$ generator matrices in a generator matrix $G_\mathcal{M} =\left[G_\mathcal{M}^{(1)}, G_\mathcal{M}^{(2)},\ldots, G_\mathcal{M}^{(\gamma_\mathcal{M})}\right] \in \mathbb{R}^{n \times m\gamma_\mathcal{M}}$, a matrix zonotope set is defined as 
    \begin{equation}
		\label{Eq:MatrixZonotope}
		\mathcal{M} =\left \langle C_\mathcal{M}, G_\mathcal{M}\right \rangle = \left\{X \in \mathbb{R}^{n \times m} \mid X=C_\mathcal{M}+\sum_{i=1}^{\gamma_\mathcal{M}} \beta^{(i)} G_\mathcal{M}^{(i)},-1 \leq \beta^{(i)} \leq 1\right\}.
    \end{equation}
\end{definition}

The zonotope set defined in Definition~\ref{Definition:ZonotopeSet} is employed for online reachable set computation in Section~\ref{Sec:4C_1}, leveraging its computational efficiency. The resulting reachable set is further approximated using the interval set described in Definition~\ref{Definition:IntervalSet}, which enables the extraction of upper and lower-bound information. These bounds are then incorporated as constraints in the optimization problem~\eqref{Eq:RNDDPCOptimizationProblem_Final} formulated in Section~\ref{Sec:4C_2}. Additionally, the matrix zonotope set introduced in Definition~\ref{Definition:MatrixZonotopeSet} is utilized to characterize the uncertainty in system matrices within the equivalent Koopman-based model~\eqref{Eq:KoopmanDynamicsRobust} in Section~\ref{Sec:4B_2}. This matrix zonotope set ensures that all possible realizations of the uncertain system matrices are encapsulated, preserving the robustness of both the model and the resulting control strategy.

\section{Data-Driven Modeling of Mixed Vehicle Platoon System}
\label{Sec:3}
This section introduces the problem statement, presents the basic parametric nonlinear model of the mixed vehicle platoon system, and discusses the data-driven modeling approach using the Koopman operator theory.

\subsection{Problem Statement}
\begin{figure}[ht]
	
	\centering
	{\includegraphics[width=8.8cm]{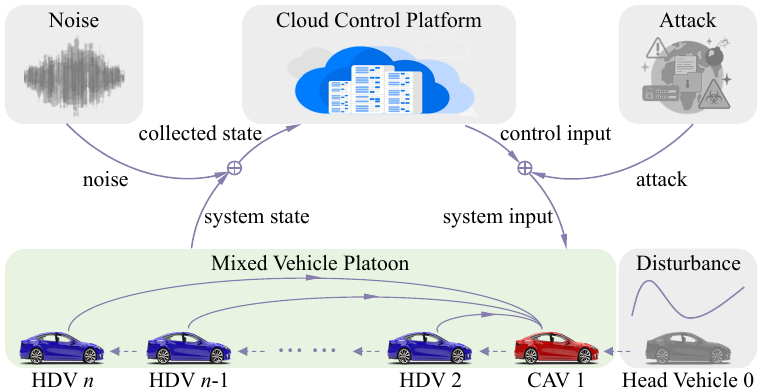}}\\
	
	\caption{The schematic of the mixed vehicle platoon consists of a leading CAV (red) and multiple following HDVs (blue), following behind a head vehicle (gray). The green box represents the mixed vehicle platoon. The disturbance originates from variations in the head vehicle's velocity. The noise and attacks affect the uplink and downlink of the cloud control platform, respectively.}
	\label{Fig:MixedPlatoon}
	\vspace{0mm}
\end{figure}

We consider a mixed vehicle platoon system consisting of one leading CAV (indexed as $ 1 $) and $ n-1$ following HDVs (indexed as $ 2, \ldots,n$), as depicted in~\figurename~\ref{Fig:MixedPlatoon}. The set of all vehicle indices in the platoon is denoted as $ \Omega =\{1, 2, \ldots, n\}$, where $ \Omega_\mathrm{C} =\{1\}$ and $\Omega_\mathrm{H} =\{2, \ldots, n\}$ represent the indices of the CAV and the HDVs, respectively. The entire platoon follows a head vehicle (indexed as $ 0 $),  which is positioned immediately ahead of the mixed vehicle platoon. This configuration is called the Car-Following Leading Cruise Control (CF-LCC) defined in~\cite{wang2021leading}.

In large-scale mixed traffic systems, the CF-LCC framework can be implemented in a decentralized manner through spatial segmentation~\cite{li2025influence}. Under this scheme, the traffic flow is partitioned into multiple mixed vehicle platoons, each consisting of a single CAV and $n-1$ HDVs, as described in~\cite{shang2023decentralized}. This decomposition offers two primary advantages. First, it reduces the original high-dimensional, multi-input system into several lower-dimensional, single-input subsystems, thereby mitigating system complexity and computational burden. Second, the decentralized architecture enhances scalability and adaptability, enabling the framework to be applied effectively across a range of traffic scales. This generality highlights the practicality of the CF-LCC framework and the rationality of using it as a representative topology for mixed vehicle platoon control.

As depicted in~\figurename~\ref{Fig:MixedPlatoon}, this study introduces a control framework designed to optimize the operation of a mixed vehicle platoon in real-time, leveraging the substantial computing capabilities of a cloud control platform. Specifically, the cloud platform collects real-time data seamlessly from all vehicles in the platoon. Based on this data, the controller generates control commands, which are transmitted to the CAV to regulate its behavior and influence the entire platoon.

Controlling mixed vehicle platoons presents several key challenges. First, the nonlinear system dynamics complicate the design of efficient and effective control strategies. Second, adverse factors, including noise in state observations, disturbances from the variation in the head vehicle's velocity, and potential adversarial attacks (e.g., false messages injected into the control channel) on the control inputs of the CAV, call for robust control. Third, the presence of potentially unknown human-in-the-loop interactions requires the CAV to effectively nudge the HDVs into adjusting their velocities in a cooperative manner, which is essential for ensuring safe operation under diverse driving conditions.

To address these challenges, this paper focuses on developing a robust nonlinear data-driven control framework that effectively addresses the complex nonlinear dynamics inherent in mixed vehicle platoons, while mitigating the impact of adverse factors. The proposed strategy aims to enhance the safety and stability of the mixed vehicle platoon, providing a novel and practical solution for real-world applications.

\subsection{Parametric Modeling for Nonlinear Dynamics}
In this part, we introduce the parametric model of the nonlinear mixed vehicle platoon system. For each vehicle $ i \in \Omega $, the longitudinal dynamics are modeled using a second-order system~\cite{wang2021leading,jin2016optimal,wang2022distributed}:
\begin{equation}
\label{Eq:DynamicsModel}
\begin{cases}
{p}_i(k+1) = {p}_i(k) + t_\mathrm{s}{v}_i(k), \\
{v}_i(k+1) = {v}_i(k) + t_\mathrm{s} u_i(k), 
\end{cases} \quad i \in \Omega
\end{equation}
where ${p}_i(k)$ and $v_i(k)$ represent the position and velocity of vehicle $ i $ at the discrete time step $ k $, $u_i(k)$ denotes the control input (acceleration), and $ t_\mathrm{s}$ is the sampling time.

For HDVs, the control input $u_i(k)$ is typically determined by the driver and is modeled using widely accepted car-following models, such as the OVM~\cite{bando1995dynamical} and IDM~\cite{treiber2000congested}. The general discrete form of these nonlinear models can be expressed as:
\begin{equation}
\label{Eq:HDV_General}
u_i(k) = F_i(s_i(k),v_i(k),v_{i-1}(k)), \quad i \in \Omega_\mathrm{H}
\end{equation}
where ${s}_i(k) = {p}_{i-1}(k) - {p}_i(k)$ denotes the spacing between vehicle $ i $ and its preceding vehicle $ i-1 $, and $F_i(\cdot)$ is a nonlinear function capturing the behavior of the HDV.

By combining~\eqref{Eq:DynamicsModel} and~\eqref{Eq:HDV_General}, and selecting $ {s}_i(k) $ and $ {v}_i(k) $ as the state variables, the state-space representation of the HDVs can be derived as:
\begin{equation}
\label{Eq:HDV_Linear}
\begin{cases}
{s}_i(k+1) = {s}_i(k) + t_\mathrm{s}\left(v_{i-1}(k) - v_i(k)\right), \\
{v}_i(k+1) = {v}_i(k) + t_\mathrm{s} F_i(s_i(k),v_i(k),v_{i-1}(k)).
\end{cases}  i \in \Omega_\mathrm{H}
\end{equation}

For CAVs, the control input is affected by adversarial attacks, similar to~\eqref{Eq:HDV_Linear}, the longitudinal dynamics of CAVs can be expressed as follows:
\begin{equation}
\label{Eq:DynamicsModelCAV_L}
\begin{cases}
{s}_i(k+1) = {s}_i(k) + t_\mathrm{s}\left(v_{i-1}(k) - v_i(k)\right), \\
{v}_i(k+1) = {v}_i(k) + t_\mathrm{s} u_i(k) + t_\mathrm{s} \vartheta_i(k), 
\end{cases} \quad i \in \Omega_\mathrm{C}
\end{equation}
where $u_i(k)$ represents the designed control input for the CAV, and $\vartheta_i(k)$ captures the non-time-delay attacks (false messages injected into the control channel) for control input. The modeling of such attack scenarios follows prior research in~\cite{xu2022reachability,khoshnevisan2025secure}.

We define the state vector for each vehicle $i$ as $x_i(k)=\left[{s}_i(k), {v}_i(k)\right]^{\top}$. By combining the state vectors of all vehicles $i \in \Omega$ in the mixed vehicle platoon, the overall system state at time step $k$ is given by: 
\begin{equation}
\label{Eq:DynamicsState}
x(k)= \left[x_1^{\top}(k),x_2^{\top}(k),\ldots,x_n^{\top}(k)\right]^{\top} \in \mathbb{R}^{2n}.
\end{equation}

Inspired by~\cite{lan2021data,zhan2022data,wang2021leading,jin2016optimal}, HDVs are treated as uncontrolled agents, while CAVs are modeled as agents under direct control. Based on~\eqref{Eq:HDV_Linear}-\eqref{Eq:DynamicsState} and accounting for the influence of noise, we formulate the nonlinear model of the mixed vehicle platoon system as follows:
\begin{equation}
\label{Eq:DynamicsSystem}
{x}(k+1)=f\left(x(k),u(k),\epsilon(k),\vartheta(k)\right)+ \omega(k),
\end{equation}
where $f(\cdot)$ denotes the nonlinear dynamics, $u(k)=u_1(k) \in \mathbb{R}$ represents the control input, $\epsilon(k)={v}_0(k) \in \mathbb{R}$ is the external disturbance (head vehicle velocity), $\vartheta(k)=\vartheta_1(k) \in \mathbb{R}$ is the adversarial attack, and $\omega(k) \in \mathbb{R}^{2n}$ is the unknown noise. The signals $\epsilon(k)$, $\vartheta(k)$, and $\omega(k)$ are assumed to satisfy the following bounded conditions: 
    \begin{equation}
		\label{Eq:all_Bound}
		\left |  \epsilon(k) \right | \leq \epsilon_\mathrm{max},\quad 
            \left |  \vartheta(k) \right | \leq \vartheta_\mathrm{max},\quad 
            \left |  \omega(k) \right  | \leq \omega_\mathrm{max},
  \end{equation}
where $\epsilon_\mathrm{max} \in \mathbb{R}$, $\vartheta_\mathrm{max} \in \mathbb{R}$, and $\omega_\mathrm{max} \in \mathbb{R}^{2n}$ denote the corresponding upper bounds.

It is important to emphasize that our proposed RNDDPC approach is entirely data-driven and does not rely on the explicit parametric nonlinear system model in~\eqref{Eq:DynamicsSystem}. The parametric nonlinear model in~\eqref{Eq:DynamicsSystem} is referenced solely to clarify the dimensions and physical significance of state and control variables, which is crucial for the design of data-driven control discussed in Section~\ref{Sec:4}.

\begin{remark}[General Form of Non–Time-Delay Attacks] 
\label{Remark_Attacks}
In this paper, the attack signal $\vartheta(k)$ in~\eqref{Eq:DynamicsSystem} represents the general form of various non–time-delay attack signals, explicitly excluding time-delay effects. Since non-time-delay attacks are typically unpredictable and difficult to characterize, we adopt a general formulation that treats them as additive signals in~\eqref{Eq:DynamicsModelCAV_L}. By refraining from assumptions regarding their precise distribution or temporal pattern, this approach provides a flexible representation capable of capturing a broad range of adversarial behaviors and thus broadens the applicability of the proposed method to diverse attack scenarios.
\end{remark}

\begin{remark}[Modeling Nonlinear Dynamics of Mixed Vehicle Platoons] 
\label{Remark_Modeling}
The inherent nonlinear dynamics of HDVs present significant challenges for accurately modeling the mixed vehicle platoon system described in~\eqref{Eq:DynamicsSystem}. Traditional model-based methods, especially those relying on linearized representations of HDVs~\cite{jin2018experimental,feng2021robust}, are fundamentally limited in capturing complex and possibly unknown HDV behaviors. 
These limitations highlight the need for a data-driven modeling approach.
To address the nonlinearity of the system dynamics, Koopman operator theory is introduced in Section~\ref{Sec:3-3} to transform nonlinear dynamics into a higher-dimensional linear representation. This transformation is achieved using state lifting function, which are obtained with the deep EDMD method proposed in Section~\ref{Sec:4B-1}, without the explicit need for knowledge about the HDV behavioral model. Overall, this Koopman operator-based approach effectively captures the system’s underlying nonlinear dynamic characteristics, providing a foundation for robust and accurate control design.
\end{remark}

\begin{remark}[Robustness Requirements in Mixed Vehicle Platoons] 
\label{Remark_Robustness}
The mixed vehicle platoon system~\eqref{Eq:DynamicsSystem} operates under the influence of noise $\omega(k)$, disturbances $\epsilon(k)$, and attacks $\vartheta(k)$. These adverse factors introduce significant challenges to maintaining robustness. However, many existing data-driven control methods fail to explicitly address these security concerns and lack robustness~\cite{wang2023deep,zhan2022data}. To overcome these limitations, we adopt a data-driven control framework based on reachability analysis, as detailed in Section~\ref{Sec:4C}. This approach systematically incorporates the effects of adverse factors into the reachable set computation. By ensuring that the reachable set remains entirely within a predefined safe region, the proposed method offers formal safety guarantees for the control of mixed vehicle platoons under adverse conditions.
\end{remark}

\subsection{Preliminary Data-Driven Modeling by Koopman Operator}
\label{Sec:3-3}
For the nonlinear system~\eqref{Eq:DynamicsSystem} with noise set to zero ($ \omega(k)=0 $), the dynamics are expressed as follows:
\begin{equation}
\label{Eq:DynamicsZeroNoise}
{x}(k+1)=f\left(x\left(k\right),u\left(k\right),\epsilon\left(k\right),\vartheta\left(k\right)\right).
\end{equation}	 
We extend the state space by taking the product of the original state space with the spaces of all control, disturbance, and attack sequences, following the rigorous and practical method in~\cite{korda2018linear}. Specifically, the extended state is defined as:
\begin{equation}
\label{Eq:Koopmanextendedstate}
\chi(k)=\left[\begin{array}{c}
x(k) \\
\boldsymbol{u}(k)
\\
\boldsymbol{\epsilon}(k)
\\
\boldsymbol{\vartheta}(k)
\end{array}\right],
\end{equation}	 
where $ \chi(k) $ denotes the extended state, $ \boldsymbol{u}(k) = [u(0),u(1),\dots,u(\infty)]^{\top}$ represents an infinite sequence of control inputs, $ \boldsymbol{\epsilon}(k)= [\epsilon(0),\epsilon(1),\dots,\epsilon(\infty)]^{\top}$ denotes an infinite sequence of disturbances, and $ \boldsymbol{\vartheta}(k) =[\vartheta(0),\vartheta(1),\dots,\vartheta(\infty)]^{\top}$ indicates an infinite sequence of attacks.

Then, the non-autonomous system dynamics~\eqref{Eq:DynamicsZeroNoise} can be reformulated into an infinite dimensional autonomous system as follows:
\begin{equation}
\label{Eq:Koopmanextendedsystem}
\chi(k+1)=\left[\begin{array}{c}
x(k+1) \\
\boldsymbol{u}(k+1)\\
\boldsymbol{\epsilon}(k+1)\\
\boldsymbol{\vartheta}(k+1)
\end{array}\right]=\left[\begin{array}{c}
f\left(x\left(k\right),{u}\left(k\right),{\epsilon}\left(k\right),{\vartheta}\left(k\right)\right) \\
\mathcal{S} \boldsymbol{u}(k)\\
\mathcal{S} \boldsymbol{\epsilon}(k)\\
\mathcal{S} \boldsymbol{\vartheta}(k)
\end{array}\right] = F\left(\chi(k)\right),
\end{equation}	 
where $F(\cdot)$ denotes the extended system dynamics, and $\mathcal{S}$ denotes the left shift operator, defined as $\mathcal{S}\boldsymbol{u}(k):= \boldsymbol{u}(k+1)$, $\mathcal{S}\boldsymbol{\epsilon}(k):= \boldsymbol{\epsilon}(k+1)$, and $\mathcal{S}\boldsymbol{\vartheta}(k):= \boldsymbol{\vartheta}(k+1)$.

It is important to note that the reformulated dynamics $\chi(k+1)=F\left(\chi(k)\right)$ in~\eqref{Eq:Koopmanextendedsystem} describe an infinite-dimensional but autonomous system. This structure allows for the direct application of the Koopman operator theory as defined in Definition~\ref{Definition:KoopmanOperator}. The Koopman operator $\mathcal{K} : \mathcal{H} \to \mathcal{H}$ enables the representation of nonlinear dynamics through the linear evolution of observables in a lifting function space (i.e., higher-dimensional space)~\cite{korda2018linear}. Specifically, the linear system evolution is described by:
\begin{equation}
\label{Eq:KoopmanOperator}
\Theta (\chi(k+1)) = \mathcal{K} \Theta (\chi(k)),
\end{equation}	
where lifting function $\Theta (\chi(k))$ belongs to the lifted space $\mathcal{H}$, which is a higher-dimensional space that transforms the original state variables, such as $x(k)$, $\boldsymbol{u}(k)$, $\boldsymbol{\epsilon}(k)$, and $\boldsymbol{\vartheta}(k)$, into new features or states.

It is important to note that the system model in~\eqref{Eq:KoopmanOperator} represents a linear operator that effectively characterizes the nonlinear dynamics of the system~\eqref{Eq:DynamicsSystem}. For mixed vehicle platoons with inherent nonlinearity (primarily due to the presence of HDVs), this linear embedding facilitates controller design using linear techniques. However, the Koopman operator $\mathcal{K}$ in~\eqref{Eq:KoopmanOperator} generally exists in an infinite-dimensional space, which arises from the infinite-dimensional nature of the variable $ \chi(k) $ in~\eqref{Eq:Koopmanextendedstate}, which presents significant computational challenges.

Inspired by the EDMD method~\cite{korda2018linear,williams2015data}, we approximate the infinite-dimensional Koopman operator $\mathcal{K}$ with a finite-dimensional operator $\mathcal{K}_{n_{\psi}} \in \mathbb{R}^{{n_{\psi}} \times {n_{\psi}}}$, where $n_{\psi}$ represents the dimension of the Koopman operator approximation. Specifically, the EDMD method involves constructing a dataset $\{[\chi(k)]_{1:n_\mathrm{c}}\}$, which is sampled independently according to a non-negative probability distribution that satisfies~\eqref{Eq:Koopmanextendedsystem}, where $n_\mathrm{c}$ denotes the number of collected data points. The finite-dimensional approximation $\mathcal{K}_{n_{\psi}}$ is then computed by solving the following optimization problem: 
\begin{equation}
\label{Eq:Koopman_KN}
\min _{\mathcal{K}_{n_{\psi}}} \sum_{j=1}^{n_\mathrm{c}-1}\left\|\Psi\left(\chi(k+1)\right)- \mathcal{K}_{n_{\psi}} \Psi\left(\chi(k)\right)\right\|_{2}^{2},
\end{equation}
where $\Psi\left(\chi(k)\right) = [\psi_{1}(\chi(k)),\psi_{2}(\chi(k)), \ldots,\psi_{n_{\psi}}(\chi(k))]^{\top}$ denotes the overall lifting function, with $\psi_{i}(\chi(k))$ representing the $i$-th basis function for $ i \in \{1,2, \ldots,n_{\psi}\} $.

To enhance the computational feasibility of the Koopman operator, the overall lifting function $\Psi\left(\chi\left(k\right)\right)$ in~\eqref{Eq:Koopman_KN} is defined as:
\begin{equation}
\label{Eq:KoopmanLiftedObservableFunctionLinear}
\Psi(\chi(k)) = \left[\begin{array}{c}
z(k) \\
u(k)\\
{\epsilon(k)}\\
{\vartheta(k)}
\end{array}\right],
\end{equation}	
where $ z(k) = [\phi_{1}(x(k)),\phi_{2}(x(k)), \ldots, \phi_{n_\mathrm{z}}(x(k))]^{\top}$ is the first block of overall lifting function $\Psi\left(\chi(k)\right)$, represents a state lifting function that project the original state $x(k)$ into a higher-dimensional feature space ($n_\mathrm{z} > 2n$), thereby enriching the representation of system dynamics. The remaining components of $\Psi(\chi(k))$ retain the original control input $u(k)$, disturbance $\epsilon(k)$, and attack $\vartheta(k)$ in explicit form. In this way, $z(k)$ captures the nonlinear features of the state, while $\Psi(\chi(k))$ aggregates both the lifted state features and the exogenous inputs into a unified representation.

Then, we represent the first $n_\mathrm{z}$ rows of $\mathcal{K}_{n_{\psi}}$ as $ \begin{bmatrix}A~B~H~J \end{bmatrix} \in \mathbb{R}^{n_\mathrm{z} \times (n_\mathrm{z} + 3)}$, where $ A \in  \mathbb{R}^{n_\mathrm{z} \times n_\mathrm{z}}$, $ B \in  \mathbb{R}^{n_\mathrm{z}}$, $ H \in  \mathbb{R}^{n_\mathrm{z}}$, and $ J \in  \mathbb{R}^{n_\mathrm{z}}$ are the system matrix, control input matrix, disturbance input matrix, and attack input matrix, respectively. The matrices $A$, $B$, $H$, and $J$ can be estimated using the data pairs $\{[x(k),u(k),\epsilon(k),\vartheta(k)]_{1:n_\mathrm{c}}\}$ by solving the least squares problem as follows~\cite{korda2018linear}:
\begin{equation}
\label{Eq:Koopman_AB}
\min _{A,B,H,J} \sum_{k=1}^{n_\mathrm{c}-1}  \left\|z(k+1) - \Xi \right\|_{2}^{2},
\end{equation}
where $\Xi = A z(k) + B u(k) + H \epsilon(k)+ J \vartheta(k)$.

To reconstruct $x(k)$ from $z(k)$, the linear mapping matrix $ C \in  \mathbb{R}^{2n \times n_\mathrm{z}}$ is obtained by solving~\cite{korda2018linear}:
\begin{equation}
\label{Eq:Koopman_C}
\min _{C} \sum_{k=1}^{n_\mathrm{c}-1}\left\|x(k+1)-C z(k+1)\right\|_{2}^{2},
\end{equation}
where $C$ is a projection matrix from the lifted space into the original state space.

By solving~\eqref{Eq:Koopman_AB} and~\eqref{Eq:Koopman_C}, the corresponding Koopman-based model for the nonlinear dynamics~\eqref{Eq:DynamicsSystem} can then be expressed as:
\begin{equation}
\label{Eq:KoopmanDynamics}
\begin{cases}
z(k+1)=A z(k)+B {u}(k)+H \epsilon(k) + J\vartheta(k), \\
{x}(k+1)=C z(k+1),
\end{cases}
\end{equation}
where $z(k)$ represents a state lifting function that project the original state $x(k)$ into a higher-dimensional feature space.

Accordingly, the nonlinear mixed vehicle platoon system~\eqref{Eq:DynamicsSystem} with zero noise ($ \omega(k)=0 $) has been approximately transformed into a high-dimensional linear system~\eqref{Eq:KoopmanDynamics} via the Koopman operator. This linearization in~\eqref{Eq:KoopmanDynamics} facilitates the formulation and solution of the mixed vehicle platoon control problem with nonlinear and possibly unknown HDV behavior. Nevertheless, as discussed in Remark~\ref{Remark_Challenges}, the Koopman-based model suffers from challenges when applied to mixed vehicle platoon control, which will be addressed in the next section.

\begin{remark}[Challenges in Koopman Operator-Based Mixed Vehicle Platoon Control] 
\label{Remark_Challenges}
The challenges in applying the Koopman operator for mixed vehicle platoon control are two-fold. First, accurately capturing the strongly nonlinear behavior of the mixed vehicle platoon, particularly due to the presence of HDVs, critically depends on the choice of the state lifting function $z(k)$ in~\eqref{Eq:KoopmanLiftedObservableFunctionLinear}. To address this, we propose a learning-based method, i.e., deep EDMD in Section~\ref{Sec:4B-1}, where a DNN is used to automatically learn a suitable $z(k)$ from data. Second, the presence of measurement noise $\omega(k)$ inevitably introduces modeling errors when replacing the original system~\eqref{Eq:DynamicsSystem} with the Koopman-based model~\eqref{Eq:KoopmanDynamics}, which can lead to safety violations. As discussed in the context of Koopman's fundamental theory in~\cite{zhang2022robust,han2023robust}, these modeling errors may degrade the performance of platoon control systems. Existing methods have not fully addressed these modeling inaccuracies~\cite{zhan2022data,korda2018linear,korda2020koopman}. To enhance robustness, we do not directly use the Koopman-based model in~\eqref{Eq:KoopmanDynamics} for controller synthesis. Instead, as discussed in Section~\ref{Sec:4B_2}, we apply matrix zonotope techniques to perform secondary learning and construct an over-approximated set of system models that capture the uncertainty arising from modeling errors. This construction serves as the foundation for the Koopman-based reachable set predictor developed in Section~\ref{Sec:4C_1}, and ultimately supports the RNDDPC strategy proposed in Section~\ref{Sec:4C_2}.
\end{remark}

\section{Robust Nonlinear Data-Driven Predictive Control for Mixed Vehicle Platoons}
\label{Sec:4}
In this section, we propose a Robust Nonlinear Data-Driven Predictive Control (RNDDPC) framework, which integrates the Koopman operator and reachability analysis, as illustrated in~\figurename~\ref{Fig:RNDDPC}. The proposed RNDDPC framework first employs a Koopman-based model to address the nonlinear modeling challenge of the mixed vehicle platoon system. To further account for modeling uncertainties, matrix zonotope sets are utilized to over-approximate the system dynamics and capture all possible system models. Based on these components, a RNDDPC optimization problem is formulated to ensure closed-loop robustness. Precisely, the RNDDPC framework is structured into three main phases:

\begin{figure*}[ht]
	\vspace{0mm}
	\centering
	{\includegraphics[width=16.5cm]{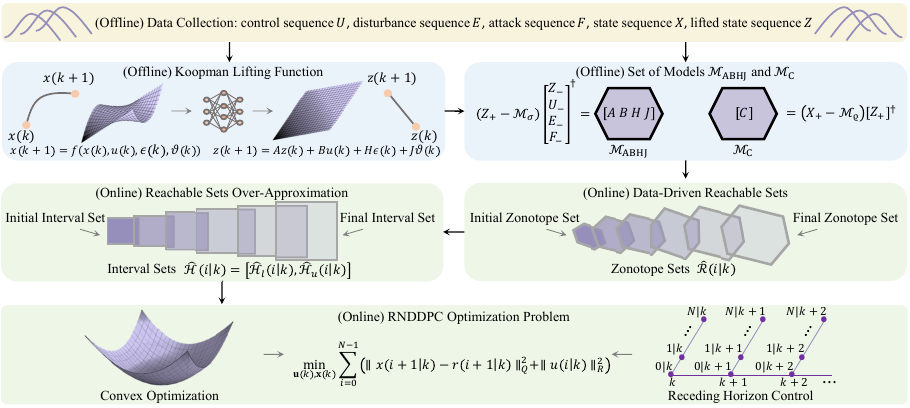}}\\
	\caption{The schematic of the proposed RNDDPC method for mixed vehicle platoons. In the offline learning phase (blue), pre-collected data (yellow) is utilized to train a DNN that learns the state lifting function $z(k)$. Using $z(k)$, this phase then computes over-approximated matrix zonotope sets $\mathcal{M}_{\mathrm{ABHJ}}$ and $\mathcal{M}_{\mathrm{C}}$, corresponding to the Koopman system matrices $\begin{bmatrix}A~B~H~J\end{bmatrix}$ and $C$, respectively. In the online control phase (green), the RNDDPC framework focuses on determining a robust optimal control input. Specifically, it utilizes the matrix zonotope sets $\mathcal{M}_{\mathrm{ABHJ}}$ and $\mathcal{M}_{\mathrm{C}}$ to perform an online recursive computation of the data-driven reachable set for the system state, represented as zonotope. To enhance computational efficiency, the zonotope-type reachable sets are approximated as interval-type reachable sets, providing straightforward upper and lower bounds for each state variable. Based on these interval bounds and predefined safety constraints, a convex optimization problem is formulated within the RNDDPC framework. This optimization problem is solved using a receding horizon control strategy, ensuring that the computed control input for the CAV maintains robust and safe operation in real-time.}
	\label{Fig:RNDDPC}
\end{figure*}

1) Data Collection Phase: This phase involves gathering data that are essential for the control framework. The dataset comprises the control inputs $u(k)$ for the CAV, the disturbance inputs ${\epsilon}(k)$ from the head vehicle, the attack inputs ${\vartheta}(k)$, and the system state $x(k)$ of the mixed vehicle platoon system. All data are collected under the influence of noise $\omega(k)$, which is, however, not directly measurable (see Section~\ref{Sec:4A}).

2) Offline Learning Phase: Using the collected data, a deep neural network (DNN) is trained to approximate the state lifting function $z(k)$. This state lifting function is then utilized to construct matrix zonotope sets, $\mathcal{M}_{\mathrm{ABHJ}}$ and $\mathcal{M}_{\mathrm{C}}$, through secondary learning, to model the uncertain dynamics of the mixed vehicle platoon system. These matrix zonotope sets serve as over-approximations of $\begin{bmatrix}A~B~H~J\end{bmatrix}$ and $C$ in the Koopman-based model~\eqref{Eq:KoopmanDynamics}, effectively capturing the uncertain dynamics of the mixed vehicle platoon. The learned $\mathcal{M}_{\mathrm{ABHJ}}$ and $\mathcal{M}_{\mathrm{C}}$ are subsequently utilized during the online control phase to calculate the data-driven reachable set (see Section~\ref{Sec:4B}).

3) Online Control Phase: Using the learned matrix zonotope sets $\mathcal{M}_{\mathrm{ABHJ}}$ and $\mathcal{M}_{\mathrm{C}}$, we recursively calculate the over-approximated data-driven reachable set of system state over the control horizon. The reachable set captures all potential states that the system can reach, accounting for adverse factors, including modeling errors, disturbances, and attacks. This ensures a robust representation of system behavior. To improve computational efficiency, the zonotope-type reachable sets are approximated as interval-type reachable sets, providing upper and lower bounds for each state. By ensuring that these interval bounds stay within predefined safety constraints, a convex optimization problem is formulated for the RNDDPC approach. This problem is solved using a receding horizon control strategy to determine the optimal control input for the CAVs, ensuring safe operation even in the presence of adverse factors (see Section~\ref{Sec:4C}).

\subsection{Data Collection Phase}
\label{Sec:4A}
In this study, the proposed RNDDPC framework is entirely data-driven. Therefore, acquiring representative data from the mixed vehicle platoon system is a prerequisite. Specifically, offline data are collected by applying control inputs (acceleration commands) to the CAVs, disturbance inputs (time-varying velocities) to the head vehicle, and attack inputs (false messages injected into the control channel) to the CAVs, thereby exciting the mixed vehicle platoon system under noisy conditions. As described by the parametric model of the mixed vehicle platoon system~\eqref{Eq:DynamicsSystem}, it is evident that the system state $ x(k) $ is influenced by the control input $ u(k) $, the disturbance inputs ${\epsilon}(k) $, the attack input ${\vartheta}(k)$, and the noise $ \omega(k) $. During the data collection process, $ u(k) $, ${\epsilon}(k) $, and ${\vartheta}(k) $ are manually specified, and the system state $ x(k) $ is directly measurable. Recall that, although the noise $ \omega(k) $ is unknown, it remains within a known bounded range. To ensure the data adequately captures the system's dynamic characteristics, persistently exciting input sequences $ u(k) $, $\epsilon(k)$, and $\vartheta(k)$ of length $ T+1 $ are applied to the system. Specifically, the control input sequence $U$, the disturbance input sequence $E$, the attack input sequence $F$, and the corresponding state sequence $X$ are defined as follows:
\begin{subequations}
	\label{Eq:XUEsequence}
	\begin{equation}
	U=[u(1),u(2),\ldots,u(T+1)] \in \mathbb{R} ^{1 \times (T+1)},
	\end{equation}
	\begin{equation}
	E=[\epsilon(1),\epsilon(2),\ldots,\epsilon(T+1)] \in \mathbb{R} ^{1 \times (T+1)},
	\end{equation}
        \begin{equation}
	F=[\vartheta(1),\vartheta(2),\ldots,\vartheta(T+1)] \in \mathbb{R} ^{1 \times (T+1)},
	\end{equation}
	\begin{equation}
	\label{Eq:Xsequence}
	X=[x(1),x(2),\ldots,x(T+1)] \in \mathbb{R} ^{2n \times (T+1)}.
	\end{equation}
\end{subequations}

Using $z(k)$ from~\eqref{Eq:KoopmanLiftedObservableFunctionLinear} and the state sequence $X$ in~\eqref{Eq:Xsequence}, we can obtain the lifted state sequence $Z$ of state $z(k) $ in the Koopman-based model~\eqref{Eq:KoopmanDynamics}:
\begin{equation}
\label{Eq:DataSequencesKoopman}
Z = [z(1),z(2),\ldots,z(T+1)] \in \mathbb{R} ^{n_\mathrm{z} \times (T+1)}.
\end{equation}

Next, all these collected or computed data are processed into standardized formats, which are then used to construct the matrix zonotope sets $\mathcal{M}_{\mathrm{ABHJ}}$ and $\mathcal{M}_{\mathrm{C}}$ for reachable set computation, as illustrated in~\figurename~\ref{Fig:RNDDPC}. Specifically, the data sequences are reorganized as follows: 
\begin{subequations}
	\label{Eq:DataSequences}
	\begin{equation}
        \label{Eq:DataSequences_A}
	U_-=[u(1),u(2),\ldots,u(T)] \in \mathbb{R} ^{1 \times T},
	\end{equation}
	\begin{equation}
        \label{Eq:DataSequences_B}
	E_-=[\epsilon(1),\epsilon(2),\ldots,\epsilon(T)] \in \mathbb{R} ^{1 \times T},
	\end{equation}
     \begin{equation}
     \label{Eq:DataSequences_C}
	F_-=[\vartheta(1),\vartheta(2),\ldots,\vartheta(T)] \in \mathbb{R} ^{1 \times T},
	\end{equation}
	\begin{equation}
       \label{Eq:DataSequences_D}
	X_-=[x(1),x(2),\ldots,x(T)] \in \mathbb{R} ^{2n \times T},
	\end{equation}
	\begin{equation}
       \label{Eq:DataSequences_E}
	X_+=[x(2),x(3),\ldots,x(T+1)] \in \mathbb{R} ^{2n \times T},
	\end{equation}
	\begin{equation}
        \label{Eq:DataSequences_F}
	Z_-=[z(1),z(2),\ldots,z(T)] \in \mathbb{R} ^{n_\mathrm{z} \times T},
	\end{equation}
	\begin{equation}
        \label{Eq:DataSequences_G}
	Z_+=[z(2),z(3),\ldots,z(T+1)] \in \mathbb{R} ^{n_\mathrm{z} \times T}.
	\end{equation}
\end{subequations}

\subsection{Offline Learning Phase}
\label{Sec:4B}
The offline learning phase seeks to address the challenge of obtaining suitable lifting function, as discussed in Remark~\ref{Remark_Challenges}, when applying the Koopman operator in mixed vehicle platoon control. Specifically, we utilize the pre-collected data to train a DNN to learn the state lifting function, which enables us to accurately approximate the nonlinear dynamics of the mixed vehicle platoon system within a higher-dimensional space. We then construct over-approximated matrix zonotope sets $\mathcal{M}_{\mathrm{ABHJ}}$ and $\mathcal{M}_{\mathrm{C}}$ through a secondary learning process based on this state lifting function, to model the nonlinear dynamics of the mixed vehicle platoon system. These matrix zonotope sets $\mathcal{M}_{\mathrm{ABHJ}}$ and $\mathcal{M}_{\mathrm{C}}$ play a critical role in the online control phase, as discussed in Section~\ref{Sec:4C}.

\subsubsection{Deep Neural Network Lifting Function} 
\label{Sec:4B-1}
\begin{figure*}[t]
	\vspace{0mm}
	\centering
	{\includegraphics[width=16.3cm]{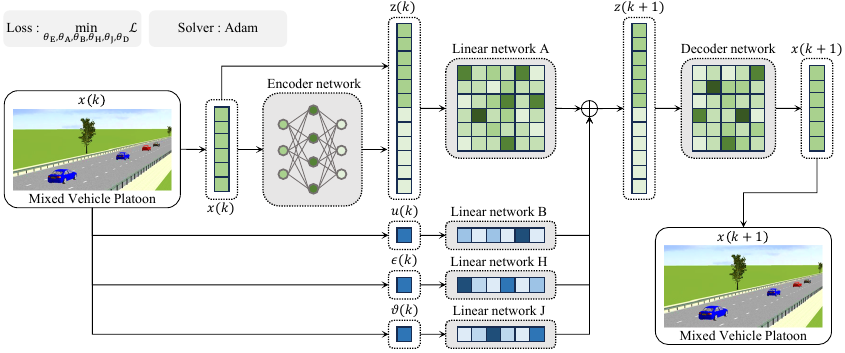}}\\
	\caption{The overview of the Koopman deep neural network framework. The original state $ x(k) \in \mathbb{R}^{2n}$ of the mixed vehicle platoon is lifted using both the state itself and an encoder network to obtain the lifted state $z(k) \in \mathbb{R}^{n_\mathrm{z}}$. The lifted state $z(k)$, along with the control input $ u(k) \in \mathbb{R}$, disturbance input $\epsilon(k) \in \mathbb{R}$, and attack input $\vartheta(k) \in \mathbb{R}$ together construct a linear evolution in the lifted space, which generates the next lifted state $z(k+1) \in \mathbb{R}^{n_\mathrm{z}}$. The corresponding original state $ x(k+1) \in \mathbb{R}^{2n}$ is then reconstructed from the lifted state space via the decoder network. The state lifting function $z(k)$ is derived through automated training of the network, guided by a well-designed loss function.}
	\label{Fig:KoopmanDNN}
\end{figure*}

To derive high-quality matrices $A$, $B$, $H$, $J$, and $C$ for the Koopman-based model~\eqref{Eq:KoopmanDynamics}, which characterizes the nonlinear dynamics in~\eqref{Eq:DynamicsSystem}, a key challenge is the selection of the appropriate state lifting function $z(k)$ in~\eqref{Eq:KoopmanLiftedObservableFunctionLinear}. While common choices include radial basis functions, polynomials functions, and Gauss functions, determining the optimal state lifting function remains an open research problem~\cite{cibulka2019data,xiao2022deep}. To tackle this challenge, we employ the deep EDMD algorithm for modeling mixed vehicle platoon systems, where DNN is employed to automatically construct the state lifting function $z(k)$.

Following the method in~\cite{xiao2022deep}, the state lifting function $z(k)$ in~\eqref{Eq:KoopmanLiftedObservableFunctionLinear} at time step $k$ are formulated as: 
\begin{equation}
\label{Eq:LiftedFunction}
z(k) =\left[\begin{array}{c}x(k)\\g(k)\end{array}\right],
\end{equation}
where $x(k)$ represents the system state as given in~\eqref{Eq:DynamicsState}, corresponding to the first $2n$ basis functions of $z(k)$, denoted by $x(k)= [\phi_{1}(x(k)),\phi_{2}(x(k)), \ldots,\phi_{2n}(x(k))]^{\top}$. The remaining $n_\mathrm{z}-2n$ basis functions, represented as $g(k) = [\phi_{2n+1}(x(k)),\phi_{2n+2}(x(k)), \ldots,\phi_{n_\mathrm{z}}(x(k))]^{\top}$, are learned by the encoder network. The architecture of the DNN is illustrated in~\figurename~\ref{Fig:KoopmanDNN}, comprising three main components: an encoder network, a linear network, and a decoder network.

\textit{Encoder Network:} The encoder network maps the system state $x(k)$ to a high-dimensional space $g(k)$, which consists of $n_\mathrm{z} - 2n$ basis functions. This network is constructed as a multi-layer perceptron with $ n_\mathrm{e} $ fully connected layers. For any hidden layer $p \in \{1, 2, \ldots, n_\mathrm{e}\}$, the output is defined as follows: 
\begin{equation}
{o}_\mathrm{e}^{(p)}=\sigma_\mathrm{e}^{(p)}\left(W_\mathrm{e}^{(p)}{o}_\mathrm{e}^{(p-1)}+b_\mathrm{e}^{(p)}\right),
\end{equation}
where $ {o}_\mathrm{e}^{(p)} $ and $ {o}_\mathrm{e}^{(p-1)} $ represent the output and input of layer $p$, respectively. The activation function for layer $p$ is $\sigma_\mathrm{e}^{(p)}$. The weight matrix $ W_\mathrm{e}^{(p)} \in \mathbb{R}^{n_{p} \times n_{p-1}}$ and the bias vector $b_\mathrm{e}^{(p)} \in \mathbb{R}^{n_{p}}$ are associated with layer $p$, where $n_{p}$ is the number of neurons in layer $p$. The encoder network takes $ {o}_\mathrm{e}^{(0)} = x(k)$ as input and produces ${o}_\mathrm{e}^{(n_\mathrm{e})}=g(k)$ as output. Rectified linear unit (ReLU) activations are applied to the layers for $p \in \{1, 2, \ldots, n_\mathrm{e}-1\}$, while the final layer does not utilize activations. By combining the state $x(k)$ with the output of encoder network $g(k)$, the lifted state $z(k)$ in~\eqref{Eq:LiftedFunction} can be obtained.

\textit{Linear Network:} The linear network consists of four sub-networks, linear network A, linear network B, linear network H, and linear network J, as shown in~\figurename~\ref{Fig:KoopmanDNN}. These sub-networks are specifically designed to approximate the matrices $ A $, $ B $, $H$, and $J$ in~\eqref{Eq:KoopmanDynamics}, respectively. The output of the linear network is expressed as:
\begin{equation}
z(k+1)=W_\mathrm{A}z(k) + W_\mathrm{B}u(k)+W_\mathrm{H}\epsilon(k)+W_\mathrm{J}\vartheta(k),
\end{equation}
where $W_\mathrm{A} \in \mathbb{R}^{n_\mathrm{z} \times n_\mathrm{z}}$, $W_\mathrm{B} \in \mathbb{R}^{n_\mathrm{z}}$, $W_\mathrm{H} \in \mathbb{R}^{n_\mathrm{z}}$, and $W_\mathrm{J} \in \mathbb{R}^{n_\mathrm{z}}$ represent the weights of linear network A, B, H, and J, respectively. To enhance the efficiency of backpropagation and accelerate convergence, the biases in all sub-networks of the linear networks are set to zero.

\textit{Decoder Network:} The decoder network approximates the matrix $C$ in~\eqref{Eq:KoopmanDynamics} and reconstructs the original system state from the high-dimensional lifted state. Similar to the linear network, the output of the decoder network is the forward-predicted state, expressed as:
\begin{equation}
{x}(k+1)=W_\mathrm{C}z(k+1),
\end{equation}
where $W_\mathrm{C}$ denotes the weights of the decoder network.

To achieve excellent performance of the DNN, we train the DNN by solving the following optimization problem:
\begin{equation}
\min_{{\theta}_\mathrm{E},{\theta}_\mathrm{A}, {\theta}_\mathrm{B}, {\theta}_\mathrm{H}, {\theta}_\mathrm{J}, {\theta}_\mathrm{D}}\mathcal{L},
\end{equation} 
where ${\theta}_\mathrm{E}$, ${\theta}_\mathrm{A}$, ${\theta}_\mathrm{B}$, ${\theta}_\mathrm{H}$, ${\theta}_\mathrm{J}$, and ${\theta}_\mathrm{D}$ denote the learning parameters for the encoder network, linear network A, linear network B, linear network H, linear network J, and decoder network, respectively. The loss function $\mathcal{L} $ for learning the DNN parameters is defined as:
\begin{equation}
\label{Eq:TotalLoss}
\mathcal{L} = \alpha_{1}\mathcal{L}_\mathrm{p} + \alpha_{2}\mathcal{L}_\mathrm{l} + \alpha_{3}\mathcal{L}_\mathrm{r} + \alpha_{4}\mathcal{L}_\mathrm{e},
\end{equation} 
where $\alpha_{1}$, $\alpha_{2}$, $\alpha_{3}$, and $\alpha_{4}$ are weighting coefficients, and $ \mathcal{L}_\mathrm{p} $, $\mathcal{L}_\mathrm{l}$, $\mathcal{L}_\mathrm{r}$, and $\mathcal{L}_\mathrm{e}$ represent the total prediction loss, linear transform loss, data reconstruction loss, and regularization loss, respectively.

\textit{Total Prediction Loss:} To ensure accurate prediction of the next step state of the original system, the total prediction loss is defined as:
\begin{equation}
\label{Eq:Loss}
\mathcal{L}_\mathrm{p}=\left\|{x}(k+1)-{x}_\mathrm{true}(k+1)\right\|_{2}^{2},
\end{equation} 
where ${x}_\mathrm{true}(k+1)$ represents the true value of the data.

\textit{Linear Transform Loss:} The linear transform loss is designed to minimize prediction errors in the lifted state space. It quantifies the difference between the predicted lifted state at the next step and the lifted state obtained from the original system's true state of the next step. This loss function is expressed as:
\begin{equation}
\mathcal{L}_\mathrm{l}=\left\|z(k+1)- z_\mathrm{true}(k+1)\right\|_{2}^{2},
\end{equation} 
where ${z}_\mathrm{true}(k+1)$ represents the lifted state obtained from the ${x}_\mathrm{true}(k+1)$.

\textit{Data Reconstruction Loss:} In order to minimize the data reconstruction error, the data reconstruction loss is expressed as: 
\begin{equation}
\label{Reconstruction}
\mathcal{L}_\mathrm{r}=\left\|{x(k+1)}-W_\mathrm{C}z(k+1) \right\|_{2}^{2}.
\end{equation}

\textit{Regularization Loss:} To prevent overfitting, a regularization term is incorporated:
\begin{equation}
\mathcal{L}_\mathrm{e}=\left\|{\theta}_\mathrm{E}\right\|_{2}^{2}.
\end{equation}

Based on the DNN structure and loss functions defined, we train the DNN using pre-collected data~\eqref{Eq:DataSequences}. In this paper, the training is performed with the Adam optimizer~\cite{kingma2014adam}. Upon completion of training, the trained DNN provides the state lifting function $z(k)$, implemented as a encoder network, along with the identified Koopman system matrices $A$, $B$, $H$, $J$, and $C$ in~\eqref{Eq:KoopmanDynamics}.

\begin{remark}[Reliability of Mapping Physical State from the Lifted State] 
\label{Remark_Reconstruction}
The projection matrix $C$ plays a critical role in mapping the lifted state $z(k)$ back to the physical state $x(k)$. To ensure that the matrix $C$ is a reliable mapping, the proposed Koopman network embeds $x(k)$ explicitly within the lifted state $z(k)$ as defined in~\eqref{Eq:LiftedFunction}. Consequently, the encoder focuses solely on learning the additional lifting function \(g(k)\), which captures the system’s nonlinear dynamics. This design reduces the complexity of the encoder and preserves the explicit representation of the physical state in the lifted space. In addition, a composite loss function~\eqref{Eq:TotalLoss} is employed during training, consisting of prediction loss, linear transform loss, data reconstruction loss, and regularization loss. In particular, the data reconstruction loss~\eqref{Reconstruction} directly penalizes mismatches between the physical state and its backward projection, thereby guiding the network to minimize projection errors. By integrating explicit state embedding with carefully designed loss terms, the proposed approach systematically enhances the reliability of mapping the physical state from the lifted state.
\end{remark}

\subsubsection{Over-Approximated System Model} 
\label{Sec:4B_2}
As highlighted in Remark~\ref{Remark_Challenges}, employing the Koopman operator to model the mixed vehicle platoon system in~\eqref{Eq:DynamicsSystem} with the linear representation in~\eqref{Eq:KoopmanDynamics} inherently introduces unavoidable modeling errors. Therefore, it is essential to develop a robust control strategy that ensures robustness concerning these errors in~\eqref{Eq:KoopmanDynamics}. To address this challenge, we first formulate an equivalent Koopman-based model of~\eqref{Eq:DynamicsSystem} that explicitly incorporates modeling errors:
\begin{equation}
\label{Eq:KoopmanDynamicsRobust}
\begin{cases}
z(k+1)=A z(k)+B {u}(k)+H \epsilon(k) + J\vartheta(k) + \sigma(k), \\
{x}(k+1)=C z(k+1) + \varrho(k),
\end{cases}
\end{equation}
where $\sigma(k)$ and $\varrho(k)$ represent the modeling errors of the Koopman-based model, arising from a combination of modeling inaccuracies and noise. Motivated by~\cite{khoshnevisan2025secure,zhang2022robust}, we assume that the uncertainty terms $\sigma(k)$ and $\varrho(k)$ are bounded by:
\begin{equation}
\label{Eq:W_Bound}
\left |  \sigma(k) \right | \leq \sigma_\mathrm{max},\quad 
\left |  \varrho(k) \right | \leq \varrho_\mathrm{max},
\end{equation}
where $\sigma_\mathrm{max} \in \mathbb{R}^{n_\mathrm{z}}$ and $\varrho_\mathrm{max} \in \mathbb{R}^{2n}$ denote the bounds for $\sigma(k)$ and $\varrho(k)$, respectively. The values of $\sigma_\mathrm{max}$ and $\varrho_\mathrm{max}$ are obtained from data by evaluating the statistical characteristics of the prediction errors generated by the Koopman-based model.

To represent the bounded modeling errors in a computationally feasible manner, we transform $\sigma(k)$ and $\varrho(k)$ in~\eqref{Eq:W_Bound} into zonotope sets. Specifically, we have:
	\begin{equation}
	\label{Eq:W_Zonotope}
	\sigma(k) \in \mathcal{Z}_{\sigma}  = \left \langle c_{\mathcal{Z}_{\sigma}}, G_{\mathcal{Z}_{\sigma}}\right \rangle = \left \langle \mathbf{0}_{n_\mathrm{z} \times 1}, \operatorname{diag}\left(\sigma_\mathrm{max}\right)\right \rangle,\quad 
	\varrho(k) \in \mathcal{Z}_{\varrho} = \left \langle c_{\mathcal{Z}_{\varrho}}, G_{\mathcal{Z}_{\varrho}}\right \rangle = \left \langle \mathbf{0}_{2n \times 1}, \operatorname{diag}\left(\varrho_\mathrm{max}\right)\right \rangle,
	\end{equation}
where $\mathcal{Z}_{\sigma}$ and $\mathcal{Z}_{\varrho}$ represent zonotope sets for $\sigma(k)$ and $\varrho(k)$, respectively. The centers are chosen as zero vectors of appropriate dimensions, while the generator matrices $G_{\mathcal{Z}{\sigma}}$ and $G_{\mathcal{Z}{\varrho}}$ are constructed from $\sigma_\mathrm{max}$ and $\varrho_\mathrm{max}$. This formulation captures the essential features of practical uncertainty while avoiding overly conservative enclosures.

For the equivalent Koopman-based model in~\eqref{Eq:KoopmanDynamicsRobust}, modeling errors may result in multiple possible representations for $\begin{bmatrix}A~B~H~J\end{bmatrix}$ and $C$ that remain consistent with the data sequences in~\eqref{Eq:DataSequences}. To account for this uncertainty, we construct matrix zonotope sets $\mathcal{M}_{\mathrm{ABHJ}}$ and $\mathcal{M}_{\mathrm{C}}$ to over-approximate all possible system models $\begin{bmatrix}A~B~H~J\end{bmatrix}$ and $C$. These $\mathcal{M}_{\mathrm{ABHJ}}$ and $\mathcal{M}_{\mathrm{C}}$ are designed to enclose every possible model that aligns with the training data, thereby ensuring that the uncertainty in the learned Koopman-based model is systematically accounted for. As a result, the robustness of the system model is preserved, which in turn enhances the reliability of the derived control strategy under uncertain conditions. The following Lemma \ref{Lemma:M_ABHJ}, extended from~\cite{lan2021data,alanwar2022robust}, establishes the approach for this over-approximation.

\begin{lemma}[Over-Approximated System Model]      
\label{Lemma:M_ABHJ}
Consider the data sequences $U_{-}$, $E_{-}$, $F_{-}$, $X_{-}$, and $X_{+}$ in~\eqref{Eq:DataSequences} obtained from the mixed vehicle platoon system~\eqref{Eq:DynamicsSystem}, along with $Z_-$ and $Z_+$ in~\eqref{Eq:DataSequences} derived from the state lifting function $z(k)$ in~\eqref{Eq:KoopmanLiftedObservableFunctionLinear} trained via a DNN. Assuming that the matrices $ \begin{bmatrix} Z_{-}^{\top}~U_{-}^{\top}~E_{-}^{\top}~F_{-}^{\top}\end{bmatrix}^{\top}$ and $ \begin{bmatrix} Z_{+} \end{bmatrix} $ are of full row rank, the sets of all possible matrices $\begin{bmatrix}A~B~H~J\end{bmatrix}$ and $\begin{bmatrix}C\end{bmatrix}$ consistent with the data can be over-approximated as follows:
	\begin{equation}
	\label{Eq:M_ABHJ}
	\mathcal{M}_{\mathrm{ABHJ}}=\left(Z_{+}-\mathcal{M}_{\sigma}\right)\begin{bmatrix}
	Z_{-} \\
	U_{-} \\
	E_{-} \\
    F_{-} 
	\end{bmatrix}^{\dagger},
	\end{equation}
	\begin{equation}
	\label{Eq:M_C}
	\mathcal{M}_{\mathrm{C}}=\left(X_{+}-\mathcal{M}_{\varrho}\right)	\begin{bmatrix}
	Z_{+}
	\end{bmatrix}^{\dagger},
	\end{equation}
	where $\dagger$ denotes the Moore–Penrose pseudoinverse. The error terms $\mathcal{M}_{\sigma}$ and $\mathcal{M}_{\varrho}$ are represented as matrix zonotopes defined by:
	\begin{equation}
	\label{Eq:setofnoise_1}
	\mathcal{M}_{\sigma} = \left \langle C_{\mathcal{M}_{\sigma}}, \left[G_{\mathcal{M}_{\sigma}}^{(1)}, G_{\mathcal{M}_{\sigma}}^{(2)},\ldots, G_{\mathcal{M}_{\sigma}}^{(n_\mathrm{z}T)}\right]\right \rangle,
	\end{equation}
    \begin{equation}
	\label{Eq:setofnoise_2}
	\mathcal{M}_{\varrho} = \left \langle C_{\mathcal{M}_{\varrho}}, \left[G_{\mathcal{M}_{\varrho}}^{(1)}, G_{\mathcal{M}_{\varrho}}^{(2)},\ldots, G_{\mathcal{M}_{\varrho}}^{(2nT)}\right]\right \rangle,
	\end{equation}
    which are constructed from the error zonotope sets $\mathcal{Z}_{\sigma} = \left \langle c_{\mathcal{Z}_{\sigma}}, G_{\mathcal{Z}_{\sigma}}\right \rangle$ and $\mathcal{Z}_{\varrho} = \left \langle c_{\mathcal{Z}_{\varrho}}, G_{\mathcal{Z}_{\varrho}}\right \rangle$ in~\eqref{Eq:W_Zonotope}. The specific constructions are given as:
	\begin{subequations}
		\begin{equation}
		C_{\mathcal{M}_{\sigma}}=\begin{bmatrix}
		c_{\mathcal{Z}_{\sigma}} & \ldots & c_{\mathcal{Z}_{\sigma}} 
		\end{bmatrix} \in \mathbb{R} ^{n_\mathrm{z} \times T},
		\end{equation}
		\begin{equation}
		G_{\mathcal{M}_{\sigma}}^{(1+(i-1) T)}=\begin{bmatrix}
		g_{\mathcal{Z}_{\sigma}}^{(i)} & \mathbf{0}_{n_\mathrm{z} \times(T-1)}
		\end{bmatrix} \in \mathbb{R} ^{n_\mathrm{z} \times T},
		\end{equation}
		\begin{equation}
		G_{\mathcal{M}_{\sigma}}^{(k+(i-1) T)}=\begin{bmatrix}
		\mathbf{0}_{n_\mathrm{z} \times(k-1)} & g_{\mathcal{Z}_{\sigma}}^{(i)} & \mathbf{0}_{n_\mathrm{z} \times(T-k)}
		\end{bmatrix} \in \mathbb{R} ^{n_\mathrm{z} \times T},
		\end{equation}
		\begin{equation}
		G_{\mathcal{M}_{\sigma}}^{(T+(i-1) T)}=\begin{bmatrix}
		\mathbf{0}_{n_\mathrm{z} \times(T-1)} & g_{\mathcal{Z}_{\sigma}}^{(i)}
		\end{bmatrix} \in \mathbb{R} ^{n_\mathrm{z} \times T},
		\end{equation}
	\end{subequations}
    and
	\begin{subequations}
		\begin{equation}
		C_{\mathcal{M}_{\varrho}}=\begin{bmatrix}
		c_{\mathcal{Z}_{\varrho}} & \ldots & c_{\mathcal{Z}_{\varrho}}
		\end{bmatrix} \in \mathbb{R} ^{2n \times T},
		\end{equation}
		\begin{equation}
		G_{\mathcal{M}_{\varrho}}^{(1+(j-1) T)}=\begin{bmatrix}
		g_{\mathcal{Z}_{\varrho}}^{(j)} & \mathbf{0}_{2n \times(T-1)}
		\end{bmatrix} \in \mathbb{R} ^{2n \times T},
		\end{equation}
		\begin{equation}
		G_{\mathcal{M}_{\varrho}}^{(k+(j-1) T)}=\begin{bmatrix}
		\mathbf{0}_{2n \times(k-1)} & g_{\mathcal{Z}_{\varrho}}^{(j)} & \mathbf{0}_{2n \times(T-k)}
		\end{bmatrix} \in \mathbb{R} ^{2n \times T},
		\end{equation}
		\begin{equation}
		G_{\mathcal{M}_{\varrho}}^{(T+(j-1) T)}=\begin{bmatrix}
		\mathbf{0}_{2n \times(T-1)} & g_{\mathcal{Z}_{\varrho}}^{(j)}
		\end{bmatrix} \in \mathbb{R} ^{2n \times T},
		\end{equation}
	\end{subequations}
	where $g_{\mathcal{Z}_{\sigma}}^{(i)}$ and $g_{\mathcal{Z}_{\varrho}}^{(j)}$ denote the $i$-th and $j$-th columns of $G_{\mathcal{Z}_{\sigma}}$ and $G_{\mathcal{Z}_{\varrho}}$, respectively, as defined in~\eqref{Eq:W_Zonotope}, with $\forall i=\left\{1, 2,\ldots, n_\mathrm{z}\right\}$, $\forall j=\left\{1, 2,\ldots, 2n\right\}$, and $\forall k=\{2, 3, \ldots, T-1\}$.
\end{lemma}
The detailed proof of Lemma~\ref{Lemma:M_ABHJ} is provided in~\ref{Appendix:Appendix A}.

\begin{remark}[Reducing the Conservatism in the Mixed Vehicle Platoon Control] 
\label{Remark_Conservatism}
Unlike the methods in~\cite{lan2021data,alanwar2022robust}, which directly construct matrix zonotope sets based on linear system models, our framework generates these sets using a Koopman-based model~\eqref{Eq:KoopmanDynamicsRobust}. This approach allows the reachable sets to be estimated in nonlinear system settings, thereby providing a more accurate characterization of system dynamics. Since the over-approximation of reachable sets is a major source of conservatism in robust control design, improving model accuracy directly mitigates this issue. In particular, the Koopman-based model reduces the prediction errors of the mixed vehicle platoon dynamics, which leads to smaller uncertainty bounds $\sigma_\mathrm{max}$ and $\varrho_\mathrm{max}$ in~\eqref{Eq:W_Bound}. As a result, the constructed zonotope sets $\mathcal{Z}_{\sigma}$ and $\mathcal{Z}_{\varrho}$ are tighter in~\eqref{Eq:W_Zonotope}, yielding less conservative safety margins.
\end{remark}

\subsection{Online Control Phase}
\label{Sec:4C}
In the online control phase, we aim to ensure the robustness of the mixed vehicle platoon control against unavoidable adverse factors. To achieve this, we develop the RNDDPC method, which incorporates the Koopman operator and reachability analysis. Initially, we calculate the reachable sets of the system's state over the control horizon, incorporating the effects of these unfavorable factors. To guarantee that the reachable sets remain within the defined safety constraints, we then formulate an optimization problem of the RNDDPC method, which we solve to determine the optimal control inputs for the CAVs.

\vspace{0.15cm}
\subsubsection{Data-Driven Reachable Set} 
\label{Sec:4C_1}
     Utilizing the model matrix zonotope sets $ \mathcal{M}_{\mathrm{ABHJ}}$ and $\mathcal{M}_{\mathrm{C}}$ derived in~\eqref{Eq:M_ABHJ} and~\eqref{Eq:M_C}, the data-driven reachable set for the state of dynamics in~\eqref{Eq:KoopmanDynamicsRobust} can be determined through Lemma~\ref{Lemma:ReachableSet}. It is important to note that this reachability analysis in Lemma~\ref{Lemma:ReachableSet} is performed using the equivalent Koopman-based model~\eqref{Eq:KoopmanDynamicsRobust}, which is tailored for nonlinear dynamics. This approach differs from existing linear-system-based methods such as those in~\cite{lan2021data,alanwar2022robust}, and thus extends data-driven reachable set to a broader class of systems.

\begin{lemma}[Data-Driven Reachable Set]        
	\label{Lemma:ReachableSet}
	Consider the system in~\eqref{Eq:KoopmanDynamicsRobust}. Given the trajectories $U_{-}$, $E_{-}$, $F_{-}$, $X_{-}$, $X_{+}$, $Z_{-}$, and $Z_{+}$, if the matrix $ \begin{bmatrix}Z_{-}^{\top}~U_{-}^{\top}~E_{-}^{\top}~F_{-}^{\top}\end{bmatrix}^{\top} $ and the matrix $ \begin{bmatrix}
	Z_{+}
	\end{bmatrix} $ are of full row rank, the recursive expression for the over-approximated data-driven reachable set is given by:
	\begin{equation}
	\label{Eq:ReachableSet}
	\begin{cases}
	\hat{\mathcal{R}}^\mathrm{z}{(i+1|k)}=\mathcal{M}_{\mathrm{ABHJ}}\left(\hat{\mathcal{R}}^\mathrm{z}{(i|k)} \times \mathcal{Z}_{u} \times \mathcal{Z}_{\epsilon} \times \mathcal{Z}_{\vartheta}\right)+\mathcal{Z}_{\sigma}, \\
	\hat{\mathcal{R}}{(i+1|k)}=\mathcal{M}_{\mathrm{C}} \hat{\mathcal{R}}^\mathrm{z}{(i+1|k)} +\mathcal{Z}_{\varrho},
	\end{cases}
	\end{equation}
	where $\hat{\mathcal{R}}{(i+1|k)}$ denotes the over-approximated reachable set for the state $ x(i+1|k)$, and $\hat{\mathcal{R}}^\mathrm{z}{(i+1|k)}$ represents the over-approximated reachable set for the lifted state $ z(i+1|k)$ of the system~\eqref{Eq:KoopmanDynamicsRobust}. The terms $\mathcal{Z}_{u} = \left \langle u(i|k), 0\right \rangle$, $\mathcal{Z}_{\epsilon} = \left \langle \epsilon(k), \epsilon_\mathrm{max}\right \rangle$, and $\mathcal{Z}_{\vartheta} = \left \langle 0, \vartheta_\mathrm{max}\right \rangle$ denote the zonotope sets for control inputs, disturbance inputs, and attack inputs, respectively, derived from~\eqref{Eq:all_Bound}.
\end{lemma}
The detailed proof of Lemma~\ref{Lemma:ReachableSet} is provided in~\ref{Appendix:Appendix B}.

\subsubsection{RNDDPC Optimization Problem} 
\label{Sec:4C_2}
This part firstly presents a well-established MPC design framework for mixed vehicle platoon systems~\eqref{Eq:DynamicsSystem}. While this overview may seem extensive, it aids in understanding our proposed RNDDPC method.

For the nonlinear mixed vehicle platoon system described in~\eqref{Eq:DynamicsSystem}, a conventional approach is to develop a nonlinear MPC strategy. The primary objective of this controller is to maintain system stability and ensure safety. The nonlinear MPC controller is formulated as follows.

\vspace{0.2em}
\noindent \textbf{Nonlinear MPC:}
\begin{subequations}
	\label{Eq:NMPCOptimizationProblem}
	\begin{align}
	&\min\limits_{\mathbf{u}(k),\mathbf{x}(k)} \ \sum_{i=0}^{N-1}\left(\|x(i+1|k)-r(i+1|k)\|_{Q}^{2}+\|u(i|k)\|_{R}^{2}\right)  \label{Eq:46a}\\
	&{s.t.}\quad \nonumber \\
	&{x}(i+1|k)=f(x(i|k),u(i|k),\epsilon(i|k),\vartheta(i|k)),   \label{Eq:46b}\\
	&x(i+1|k) \in \mathcal{X},  \label{Eq:46c}\\
	&u(i|k) \in \mathcal{U}, \label{Eq:46d}\\
        &\epsilon(i|k) =\epsilon(k),  \label{Eq:46e}\\
	&x(0|k)=x(k),  \label{Eq:46f}
	\end{align}
\end{subequations}
where~\eqref{Eq:46a} represents the cost function, with the optimized variables $ \mathbf{u}(k) = \{u(0|k), u(1|k), \ldots, u(N-1|k) \}$ and $ \mathbf{x}(k) = \{x(1|k), x(2|k), \ldots, x(N|k) \}$ denoting the sequences of control input and predicted state, respectively. The term $r(i+1|k) = \left[r_1^{\top}(k),r_2^{\top}(k),\ldots,r_n^{\top}(k)\right]^{\top} \in \mathbb{R}^{2n}$, with $r_i(k)=\left[{s}^{*}_i(k), {v}^{*}_i(k)\right]^{\top}$ is the desired state to be tracked, and $ N $ is the length of the prediction horizon. The weight matrices $Q = \operatorname{diag}\left(Q_x, \xi Q_x, \ldots, \xi^{(n-1)} Q_x\right) \in \mathbb{R}^{2n \times 2n}$ and $R \in \mathbb{R}$ penalize deviations in state and control input, respectively. Specifically, $Q_x = \operatorname{diag}\left(\rho_{s}, \rho_{v}\right) $ represents the penalty weights for spacing and velocity deviations, with $\rho_{s} $ and $\rho_{v} $ are the corresponding weights. The decay factor $ 0 < \xi \le 1 $ ensures that penalties decrease for HDVs that are further from the CAV.

The~\eqref{Eq:46b} represents the nonlinear dynamics constraints from~\eqref{Eq:DynamicsSystem}, assuming that attack $\vartheta(i|k)$ and noise $\omega(i|k)$ are zero for online control. The constraint~\eqref{Eq:46c} imposes limits on the system state, where the state constraint $\mathcal{X}=\left\{x(k) \in \mathbb{R}^{2n}\mid|x(k)-r(k)|\leq \mathbf{1}_{n} \otimes \tilde{x}_{\max }\right\}$. Here, $\tilde{x}_{\max }=\left[\tilde{s}_{\max }, \tilde{v}_{\max }\right]^{\top} $, where $\tilde{s}_{\max }$ and $\tilde{v}_{\max }$ are the constraint limits for spacing and velocity deviations, respectively. Additionally, $\mathbf{1}_{n}$ is an $n$ dimensional column vector of ones and $\otimes$ denotes the Kronecker product. The~\eqref{Eq:46d} is the control input constraint, we define $\mathcal{U}=\left\{u(k) \in \mathbb{R}\mid|u(k)|\leq u_{\max }\right\}$, where $u_{\max }$ defines the maximum control input allowable for the CAVs. The~\eqref{Eq:46e} assumes that the head vehicle is running at a constant velocity $\epsilon(k)$. Finally, the optimization problem is initialized with current state measurement $x(k)$, as shown in constraint~\eqref{Eq:46f}.

Notably, the nonlinear dynamics in~\eqref{Eq:46b} can be derived through system identification methods or first-principles modeling. 
However, accurately modeling the mixed vehicle platoon systems~\eqref{Eq:DynamicsSystem}, which often involve multiple HDVs, poses significant challenges due to their strong nonlinearities. Consequently, the nonlinear MPC optimization problem in~\eqref{Eq:NMPCOptimizationProblem} is typically strongly non-convex. Solving these non-convex problems in real time requires substantial computational resources, which can potentially compromise the performance of the nonlinear MPC approach~\cite{diehl2009efficient}. Moreover, the presence of multiple local minima can undermine control effectiveness, leading to suboptimal strategies. Although linearizing~\eqref{Eq:46b} and employing a linear MPC controller may seem like a viable option, this approach may not adequately capture the inherently nonlinear dynamics of mixed vehicle platoons.

To address the nonlinear challenges inherent in nonlinear MPC framework~\eqref{Eq:NMPCOptimizationProblem}, a promising solution is the Koopman MPC method~\cite{korda2018linear,korda2020koopman}. This approach replaces the nonlinear dynamics constraints in~\eqref{Eq:46b} with high-dimensional linear dynamics as described by~\eqref{Eq:KoopmanDynamics}. In the Koopman MPC framework, the nonlinear MPC problem~\eqref{Eq:NMPCOptimizationProblem} is reformulated as a quadratic optimization problem.

\vspace{0.2em}
\noindent \textbf{Koopman MPC:}
\begin{subequations}
	\label{Eq:KoopmanOptimizationProblem}
	\begin{align}
	&\min\limits_{\mathbf{u}(k),\mathbf{x}(k)} \ \sum_{i=0}^{N-1}\left(\|x(i+1|k)-r(i+1|k)\|_{Q}^{2}+\|u(i|k)\|_{R}^{2}\right)  \label{Eq:47a}\\
	&{s.t.}\quad \nonumber \\
	&z(i+1|k)=A z(i|k)+B {u}(i|k) + H {\epsilon}(i|k) + J\vartheta(i|k),  \label{Eq:47b}\\
	&{x}(i+1|k)=C z(i+1|k), \label{Eq:47c}\\
	&x(i+1|k) \in \mathcal{X},  \label{Eq:47d}\\
	&u(i|k) \in \mathcal{U}, \label{Eq:47e}\\
        &\epsilon(i|k) =\epsilon(k), \label{Eq:47f}\\
	&x(0|k)=x(k), \label{Eq:47g}
	\end{align}
\end{subequations}
where the cost function in~\eqref{Eq:47a} is identical to~\eqref{Eq:46a}, Similarly, the safety constraints in~\eqref{Eq:47d} and~\eqref{Eq:47e} correspond to those in~\eqref{Eq:46c} and~\eqref{Eq:46d}, respectively. However, the dynamics constraints~\eqref{Eq:47b} and~\eqref{Eq:47c} derived from the linear Koopman-based model~\eqref{Eq:KoopmanDynamics}, rather than the nonlinear model in~\eqref{Eq:46b}. Same as in~\eqref{Eq:NMPCOptimizationProblem}, the attack $\vartheta(i|k)$ and noise $\omega(i|k)$ are assumed to be zero for online control.

For mixed vehicle platoon systems, however, offline data is often corrupted by unknown noise, complicating accurate modeling with the Koopman-based model~\eqref{Eq:47b} and~\eqref{Eq:47c}. To address this issue, this paper reformulates the problem as a RNDDPC problem using~\eqref{Eq:ReachableSet}. The key idea is to determine the control input $\mathbf{u}(k)$ at each time step $ k $ to ensure that the predicted state $\mathbf{x}(k)$ remain within the computed reachable sets, while also keeping these sets within the safety constraints, all while minimizing the associated cost. The RNDDPC problem can be formulated as follows.

\vspace{0.2em}
\noindent \textbf{RNDDPC:}
\begin{subequations}
	\label{Eq:RNDDPCOptimizationProblem}
	\begin{align}
	&\min\limits_{\mathbf{u}(k),\mathbf{x}(k)} \ \sum_{i=0}^{N-1}\left(\|x(i+1|k)-r(i+1|k)\|_{Q}^{2}+\|u(i|k)\|_{R}^{2}\right)  \label{Eq:48a}\\
	&{s.t.}\quad \nonumber \\
	&\hat{\mathcal{R}}^\mathrm{z}{(i+1|k)}=\mathcal{M}_{\mathrm{ABHJ}}\left(\hat{\mathcal{R}}^\mathrm{z}{(i|k)} \times \mathcal{Z}_{u} \times \mathcal{Z}_{\epsilon} \times \mathcal{Z}_{\vartheta} \right)+\mathcal{Z}_{\sigma},   \label{Eq:48b}\\
	&\hat{\mathcal{R}}{(i+1|k)}=\mathcal{M}_{\mathrm{C}} \hat{\mathcal{R}}^\mathrm{z}{(i+1|k)} +\mathcal{Z}_{\varrho},  \label{Eq:48c}\\
	&\hat{\mathcal{R}}{(i+1|k)} \subset \mathcal{X},  \label{Eq:48d}\\
	&x(i+1|k) \in \hat{\mathcal{R}}{(i+1|k)},  \label{Eq:48e}\\
	&u(i|k) \in \mathcal{U},  \label{Eq:48f}\\
	&x(0|k)=x(k),  \label{Eq:48g}\\
        &\hat{\mathcal{R}}{(0|k)}=\left \langle x(k), 0\right \rangle. &\tag{48h} \label{Eq:48h}
	\end{align}
\end{subequations}
Here, constraints~\eqref{Eq:48b} and~\eqref{Eq:48c} are derived from~\eqref{Eq:ReachableSet}. Constraint~\eqref{Eq:48d} ensures that the predicted reachable set $\hat{\mathcal{R}}{(i+1|k)}$ satisfies the allowable state constraint $\mathcal{X}$, while constraint~\eqref{Eq:48e} guarantees that the predicted state $ x(i+1|k) $ remains within the reachable set. Note that the optimization problem~\eqref{Eq:RNDDPCOptimizationProblem} is convex. The recursive feasibility and closed-loop constraint satisfaction for the RNDDPC optimization problem~\eqref{Eq:RNDDPCOptimizationProblem} is provided in~\ref{Appendix:Appendix C} and~\ref{Appendix:Appendix D}.

To implement constraint~\eqref{Eq:48d} and~\eqref{Eq:48e}, we need to confirm that the predicted reachable set $\hat{\mathcal{R}}{(i+1|k)}$ is a subset of the set $\mathcal{X}$. To achieve this, we first apply the over-approximation operation in Definition~\ref{Definition:ZonotopeSet} to approximate  $\hat{\mathcal{R}}{(i+1|k)}$ as an interval set, as outlined in Definition~\ref{Definition:IntervalSet}:
\begin{equation}
\label{Eq:Interval}
\hat{\mathcal{H}}{(i+1|k)} = \mathsf{interval}(\hat{\mathcal{R}}{(i+1|k)}) = [\hat{\mathcal{H}}_l{(i+1|k)},\hat{\mathcal{H}}_u{(i+1|k)}],
\end{equation}
where $\hat{\mathcal{H}}_l{(i+1|k)}$ and $\hat{\mathcal{H}}_u{(i+1|k)}$ are defined in~\eqref{Eq:over-approximation}.

Then, we reformulate~\eqref{Eq:RNDDPCOptimizationProblem} to obtain the executable RNDDPC problem as follows.

\noindent \textbf{Executable RNDDPC:}
\begin{subequations}
	\label{Eq:RNDDPCOptimizationProblem_Final}
	\begin{align}
	&\min\limits_{\mathbf{u}(k),\mathbf{x}(k)} \ \sum_{i=0}^{N-1}\left(\|x(i+1|k)-r(i+1|k)\|_{Q}^{2}+\|u(i|k)\|_{R}^{2}\right)  \label{Eq:50a}\\
	&{s.t.}\quad \nonumber \\
	&\hat{\mathcal{R}}^\mathrm{z}{(i+1|k)}=\mathcal{M}_{\mathrm{ABHJ}}\left(\hat{\mathcal{R}}^\mathrm{z}{(i|k)} \times \mathcal{Z}_{u} \times \mathcal{Z}_{\epsilon} \times \mathcal{Z}_{\vartheta}\right)+\mathcal{Z}_{\sigma},   \label{Eq:50b}\\
	&\hat{\mathcal{R}}{(i+1|k)}=\mathcal{M}_{\mathrm{C}} \hat{\mathcal{R}}^\mathrm{z}{(i+1|k)} +\mathcal{Z}_{\varrho},  \label{Eq:50c}\\
	&\hat{\mathcal{H}}_l{(i+1|k)} \geq \mathcal{X}_l, \quad \hat{\mathcal{H}}_u{(i+1|k)} \leq \mathcal{X}_u,  \label{Eq:50d}\\
	&x(i+1|k) \geq \hat{\mathcal{H}}_l{(i+1|k)}, \quad x(i+1|k) \leq \hat{\mathcal{H}}_u{(i+1|k)},  \label{Eq:50e}\\
	&u(i|k) \in \mathcal{U}, \label{Eq:50f}\\
	&x(0|k)=x(k), \label{Eq:50g}\\
        &\hat{\mathcal{R}}{(0|k)}=\left \langle x(k), 0\right \rangle, &\tag{50h} \label{Eq:50h}
	\end{align}
\end{subequations}
where $\mathcal{X}_l$ and $ \mathcal{X}_u $ represent the vectors of the lower and upper bounds for each state.

At each time step $k$, we solve the final RNDDPC optimization problem~\eqref{Eq:RNDDPCOptimizationProblem_Final} online using a receding horizon manner. This yields the optimal control input sequence $ \mathbf{u}^{*}(k)=[u^{*}(0|k), u^{*}(1|k), \dots,u^{*}(N-1|k)] $ and predicted state sequence $ \mathbf{x}^{*}(k)=[x^{*}(1|k), x^{*}(2|k),\dots,x^{*}(N|k)] $. The first control input $ u^{*}(0|k)$ is then applied to the CAV in the mixed vehicle platoon. The detailed procedure for the RNDDPC implementation is provided in Algorithm~\ref{algorithm1}.

\begin{algorithm}[ht]
    \caption{RNDDPC}
    \label{algorithm1}
    \SetAlgoLined
    \KwIn{Pre-collected data ($U$, $E$, $F$, $X$), system constraints ($\mathcal{X}$, $\mathcal{U}$), weighting matrices ($Q$, $R$), bounded sets ($\mathcal{Z}_{\epsilon}$, $\mathcal{Z}_{\vartheta}$, $\mathcal{Z}_{\omega}$), control horizon $N$, and total steps $N_\mathrm{s}$.}

    Offline construct data $(U_-, E_-, F_-, X_-, X_+, Z_-, Z_+)$\;

    Offline learn the state lifting function $z(k)$ using DNN\;
    
    Offline calculate $\mathcal{Z}_{\sigma}$ and $\mathcal{Z}_{\varrho}$ by~\eqref{Eq:W_Bound} and~\eqref{Eq:W_Zonotope}, and learn the matrix zonotope sets $\mathcal{M}_{\mathrm{ABHJ}}$ and $\mathcal{M}_{\mathrm{C}}$ using~\eqref{Eq:M_ABHJ} and~\eqref{Eq:M_C}\;
    Initialize the state $x(0)$ at time step $0$\;
    
    \While{$ 0 \leq k \leq N_\mathrm{s} $}{
        
        Solve~\eqref{Eq:RNDDPCOptimizationProblem_Final} for the optimal control input sequence: $\mathbf{u}^{*}(k)=[u^{*}(0|k), u^{*}(1|k), \dots,u^{*}(N-1|k)] $\; 

        Apply the first control input $ u^{*}(0|k)$ to the CAV\;

        Increase the time step $ k \gets k+1 $ and update the initial state $x(k)$ of the mixed vehicle platoon\;
    }
\end{algorithm}

\begin{remark}[Enhancing Generalization Capability via Koopman-Based Modeling and Robust Control]
\label{Remark_Generalization}
The generalization capability of the proposed RNDDPC framework is achieved through complementary advances in both modeling and control design. From the modeling perspective, the Koopman-based model integrates physically interpretable feature lifting and well-designed regularization within a lightweight structure, effectively mitigating overfitting and providing a reliable approximation of mixed vehicle platoon dynamics under diverse operating conditions. From the control perspective, generalization is further strengthened by incorporating reachability analysis into the predictive control design, which explicitly addresses system uncertainties and reduces the impact of prediction errors in unseen or rapidly changing scenarios. Together, the framework forms a hybrid architecture that unites data-driven learning with robust control theory, yielding superior adaptability and reliability across varied traffic environments.
\end{remark}

\section{Simulation and Results}
\label{Sec:5}	
In this section, we perform nonlinear simulations to assess the performance of the proposed RNDDPC method in controlling mixed vehicle platoon systems. 

\subsection{Simulation Setup}
\begin{figure}[ht]
	\centering
	{\includegraphics[width=15cm]{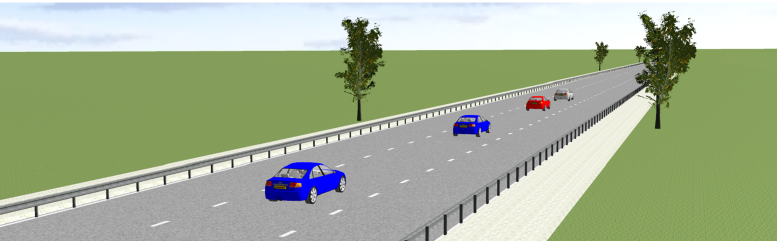}}\\
	\caption{Simulation scenario for mixed vehicle platoon control in PreScan simulation software. The CAV is highlighted in red, the HDVs in blue, and the head vehicle in gray, consistent with the representation in~\figurename~\ref{Fig:MixedPlatoon}.}
	\label{Fig:ExperimentPlatform}
\end{figure}

As shown in~\figurename~\ref{Fig:ExperimentPlatform}, the simulation setup considers $n=3$, representing a platoon composed of one CAV (red) and two HDVs (blue), with a head vehicle (gray) positioned ahead of the platoon. Unlike the simplified point-mass models widely adopted in studies such as~\cite{wang2023deep,lan2021data,zhan2022data}, our simulations employ vehicles with full two-dimensional dynamics, specifically the Audi A8 model implemented in the PreScan software~\cite{ortega2020overtaking}. This modeling choice provides a more accurate representation of vehicle behavior, improves the realism of the simulation data, and ensures that the evaluation results closely reflect real-world driving conditions.

For the HDVs ($i \in \Omega_\mathrm{H}$) in the simulation, control inputs are modeled using the OVM~\cite{bando1995dynamical}, which is expressed as: 
\begin{equation}
\label{Eq:HDVsU}
u_i(k) = k_{1_i}\left(V\left(s_i(k)\right)-v_i(k)\right)+k_{2_i}\left(v_{i-1}(k)-v_i(k)\right),
\end{equation}
where $k_{1_i}$ and $k_{2_i}$ are control parameters of the HDVs, and the desired velocity $V\left(s_{i}(k)\right)$ is defined as:
\begin{equation}
\label{Eq:HDVsModel1}
V\left(s_{i}(k)\right)=
\begin{cases}
0, & s_{i}(k) \leq s_{\min }\\
V_\mathrm{m}\left(s_{i}(k)\right), & s_{\min }<s_{i}(k)<s_{\max }\\
v_{\max }, & s_{i}(k) \geq s_{\max },
\end{cases}
\end{equation}
where $s_{\min }$ and $s_{\max }$ represent the minimum and maximum spacings, and $v_{\max }$ is the maximum velocity. The nonlinear desired velocity function $V_\mathrm{m}\left(s_{i}(k)\right)$ is defined as in~\cite{jin2018experimental}:
\begin{equation}
\label{Eq:HDVsModel2}
V_\mathrm{m}\left(s_{i}(k)\right)=\frac{v_{\max }}{2}\left(1-\cos \left(\pi \frac{s_{i}(k)-s_{\min }}{s_{\max }-s_{\min }}\right)\right).
\end{equation}

In the simulation, the parameters for the HDVs, indexed by $i \in \Omega_\mathrm{H}$ in~\eqref{Eq:HDVsU}-\eqref{Eq:HDVsModel2}, are set as follows: $k_{1_i}=0.6$, $k_{2_i}=0.9$, $v_{\max} = \SI{38}{m/s}$, $s_{\max} = \SI{35}{m}$, and $s_{\min} = \SI{5}{m}$.

For the CAV ($i \in \Omega_\mathrm{C}$) in the simulation, its control input is obtained using the proposed RNDDPC method, with the following simulation parameters:
        \begin{itemize}
	\item 
        For the data collection phase, datasets are collected near a traffic velocity of $ v^{*}=\SI{19}{m/s} $ and are utilized for all subsequent experiments. During this phase, the control input for the CAV employs the OVM model in~\eqref{Eq:HDVsU} as a pre-designed controller to facilitate normal driving operations. The head vehicle's velocity is an external disturbance input, modeled as a random fluctuation $\epsilon(k) \sim \mathbb{U}\left[17, 21\right] $, where $\mathbb{U}$ denotes the uniform distribution. Additionally, random attack inputs $\vartheta(k) \sim \mathbb{U}\left[-2, 2\right] $ and random noise $\omega(k)\sim \mathbf{1}_{2n} \otimes \mathbb{U}\left[-0.03, 0.03\right]$ are introduced. To ensure adequate persistent excitation, the offline collected trajectories have a length of $ T = 10000 $ with a sampling interval of $\SI{0.05}{s}$. These trajectories are used to construct the data sequences in~\eqref{Eq:DataSequences}.
  
        \item 
        For the offline learning phase, the collected data is split into training, testing, and validation sets in a $7:2:1$ ratio for training the DNN that estimates the state lifting function. The DNN structure is depicted in~\figurename~\ref{Fig:KoopmanDNN}, with the encoder network comprising $n_\mathrm{e}=5 $ layers with $32$, $32$, $64$, $32$, and $6$ neurons, respectively, resulting in $n_\mathrm{z}=12$ state lifting function. The linear networks for matrices $A$, $B$, $H$, and $J$ consist of layers with $12$ neurons each. The decoder network is configured with $6$ neurons. Training is performed with a batch size of $128$, a learning rate of $0.001$, and loss function weights of $\alpha_{1}=1$, $\alpha_{2}=10$, $\alpha_{3}=3$, and $\alpha_{4}=0.0001$. The model error bounds are set as $\sigma_\mathrm{max}=\mathbf{1}_{n_\mathrm{z}} \otimes 0.3$ and $\varrho_\mathrm{max}=\mathbf{1}_{2n} \otimes 0.3$ in~\eqref{Eq:W_Bound}, with a detailed procedure for extracting these uncertainty parameters provided in~\ref{Appendix:Appendix E}. Based on these specifications, the error zonotopes in~\eqref{Eq:W_Zonotope} are constructed as $\mathcal{Z}_{\varrho} = \left \langle \mathbf{0}_{2n \times 1}, \operatorname{diag}\left(\mathbf{1}_{2n} \otimes 0.3\right)\right \rangle$ and $\mathcal{Z}_{\sigma} = \left \langle \mathbf{0}_{n_\mathrm{z} \times 1}, \operatorname{diag}\left(\mathbf{1}_{n_\mathrm{z}} \otimes 0.3\right)\right \rangle$. The over-approximated system matrix zonotope sets $\mathcal{M}_{\mathrm{ABHJ}}$ and $\mathcal{M}_{\mathrm{C}}$ are then learned based on~\eqref{Eq:M_ABHJ} and~\eqref{Eq:M_C}. Their over-approximation tightness is rigorously evaluated in~\ref{Appendix:Appendix F}, providing a quantitative assessment of conservativeness.

	\item 
        For the online control phase, the simulation is performed under attack inputs $\vartheta(k) \sim \mathbb{U}\left[-3, 3\right] $ and noise $\omega(k)\sim \mathbf{1}_{2n} \otimes \mathbb{U}\left[-0.03, 0.03\right]$. The prediction horizon is set to $ N=5$. The cost function weights in~\eqref{Eq:50a} are defined as $\rho_{s} = 1$, $\rho_{v} = 1$, $ \xi=0.6 $, and $R = 0.1$. State constraints are set as $\tilde{x}_{\max } = \left[7, 7\right]^{\top}$ for comprehensive simulations and emergency simulations. Input constraint is set as $u_{\max} = 5$ in~\eqref{Eq:RNDDPCOptimizationProblem_Final}. The head vehicle’s velocity ${v}_0(k) $ is set to the desired reference velocity ${v}^{*}_{i}(k)$ for each vehicle. For the HDVs, the desired spacing ${s}^{*}_{i}(k)$ is determined by the inverse model $V^{-1}\left(s_{i}(k)\right)$ of~\eqref{Eq:HDVsModel1}. For the CAV, a constant time headway strategy is adopted, where ${s}^{*}_i(k) = t_\mathrm{h} \times {v}^{*}_{i}(k) + s_{\min}$, with a time headway of $t_\mathrm{h} = \SI{1.2}{s}$. The disturbance boundary is set to $\epsilon_\mathrm{max} = 2$, while the attack boundary is set to $\vartheta_\mathrm{max} = 3$. The simulation operates with a time step of $\SI{0.05}{s}$.
	\end{itemize}

The simulations are implemented using MATLAB 2024a and PreScan 8.5. Optimization problems are formulated with the YALMIP toolbox~\cite{lofberg2004yalmip} and solved using the Mosek solver~\cite{aps2019mosek}. Reachable sets are computed with the CORA toolbox~\cite{althoff2021guaranteed}. All simulations are conducted on a computer equipped with an Intel Core i9-13900KF CPU and 64 GB of RAM.

For comparative analysis, some baseline methods are included: 
\begin{enumerate}
    \item The nonlinear MPC (NMPC) method, assumes full knowledge of the dynamics of HDVs based on~\eqref{Eq:HDVsU}, but does not account for adverse factors, as outlined in~\eqref{Eq:NMPCOptimizationProblem}. 
    \item The linear MPC (LMPC) method linearizes the known system dynamics~\eqref{Eq:DynamicsSystem} while maintaining the same setup as the nonlinear MPC. 
    \item The Koopman MPC (KMPC) method, as described in~\cite{korda2018linear,korda2020koopman}, utilizes the same state lifting function as our proposed method and follows the formulation in~\eqref{Eq:KoopmanOptimizationProblem}. 
    \item The DeePC method is implemented based on~\cite{wang2023deep,coulson2019data}, which does not explicitly consider adverse factors.
    \item The ZPC method from~\cite{lan2021data,alanwar2022robust} employs zonotope sets for data-driven control, explicitly incorporating adverse factors. The ZPC setup aligns closely with our method, with the primary distinction being its reliance on a linear system assumption. 
\end{enumerate}

All baseline methods share identical parameter settings for the cost function~\eqref{Eq:46a} and the safety constraints~\eqref{Eq:46d} and~\eqref{Eq:46f} with RNDDPC, except that ZPC uses a prediction horizon of $N = 5$, while the other methods use $N = 10$. For further contrast, an all-HDV scenario is also included, where the CAV is controlled using the OVM in~\eqref{Eq:HDVsU}.

After completing the experiments with a total duration of $T_\mathrm{s}$, data from each vehicle is collected for performance evaluation. To quantify the tracking accuracy of the proposed RNDDPC method and the baseline approaches, two performance indices are employed: tracking velocity mean error $ R_\mathrm{v} $ and tracking spacing mean error $ R_\mathrm{s} $, which are defined as follows:
\begin{equation}
\label{Eq:VMADvalue}
R_\mathrm{v}=\frac{1}{N_\mathrm{s}} \frac{1}{n} \sum_{k=1}^{N_\mathrm{s}} \sum_{i=1}^{n}\left|v_{i}(k)-v^{*}(k)\right|,
\end{equation}
\begin{equation}
\label{Eq:SMADvalue}
R_\mathrm{s}=\frac{1}{N_\mathrm{s}} \frac{1}{n} \sum_{k=1}^{N_\mathrm{s}} \sum_{i=1}^{n}\left|s_{i}(k)-s^{*}_{i}(k)\right|,
\end{equation}	
where $ N_\mathrm{s} = {T_\mathrm{s}} \big / {t_{s}} $ represents the total number of simulation steps. In~\eqref{Eq:VMADvalue} and~\eqref{Eq:SMADvalue}, lower values of $ R_\mathrm{v} $ and $ R_\mathrm{s} $ indicate better tracking performance of the vehicles in the mixed vehicle platoon relative to the head vehicle, while higher values signify poorer tracking accuracy.

In addition, based on~\eqref{Eq:50a}, the average real cost $ R_\mathrm{c} $ is employed to evaluate the control performance of different methods, as defined by:
\begin{equation}
\label{Eq:COSTvalue}	
R_\mathrm{c}=\frac{1}{N_\mathrm{s}} \sum_{k=1}^{N_\mathrm{s}}\left(\|x(k)-r(k)\|_{Q}^{2}+\|u(k)\|_{R}^{2}\right),
\end{equation}		
where $R_\mathrm{c}$ reflects the overall control efficiency, with smaller values of $R_\mathrm{c}$ signifying more efficient control performance.

\subsection{Comprehensive Simulations}
\label{Sec:5.2}	
\begin{figure*}[ht]
	\vspace{0mm}
	\centering
	\subfigure[Comprehensive condition]{\includegraphics[width=8.1cm]{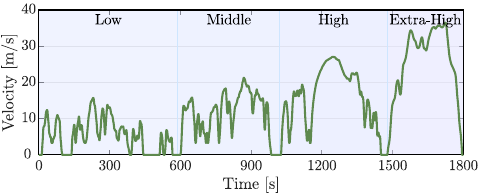}}
        \subfigure[Emergency condition]{\includegraphics[width=8.1cm]{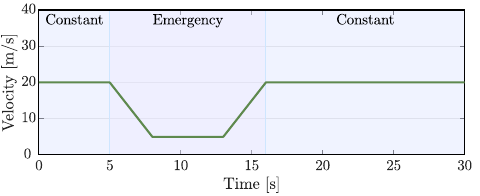}}
	\caption{The velocity profiles for comprehensive simulations and emergency simulations.}
	\label{Fig:WLTP}
\end{figure*}

Standard driving cycles are commonly employed to evaluate the performance of mixed vehicle platoon control algorithms~\cite{wang2023deep,lan2021data}. In order to evaluate the effectiveness of our method, we employ the Worldwide harmonised Light vehicles Test Procedure (WLTP) as the simulation condition~\cite{tutuianu2015development}. The velocity profile of WLTP is shown in~\figurename~\ref{Fig:WLTP}(a), which includes low velocity, middle velocity, high velocity, and extra-high velocity phases, covering a variety of driving conditions such as urban, suburban, and highway scenarios.

\begin{figure*}[!ht]
	\vspace{-2mm}
	\centering
	\subfigure[All-HDVs]{\includegraphics[width=8.1cm]{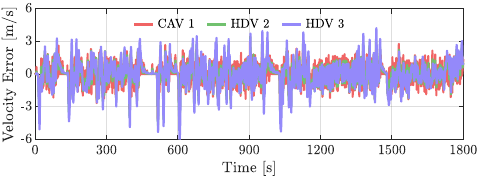} \hspace{0.5mm}
                         \includegraphics[width=8.1cm]{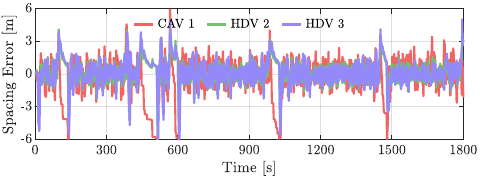}}\\
    \vspace{-1.3mm}
	\subfigure[Linear MPC]{\includegraphics[width=8.1cm]{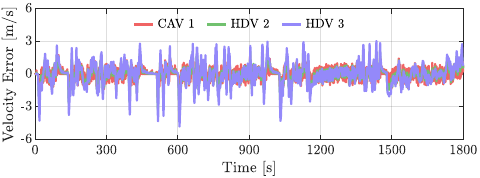} \hspace{0.5mm}
                           \includegraphics[width=8.1cm]{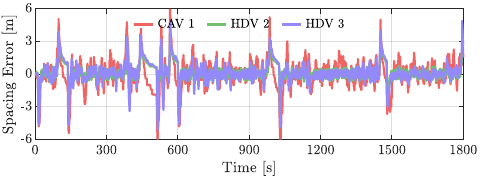}}\\
    \vspace{-1.3mm}
    \subfigure[DeePC]{\includegraphics[width=8.1cm]{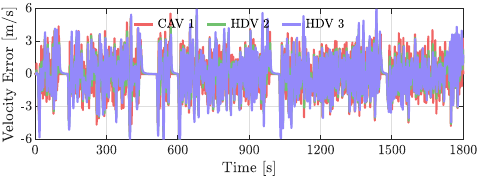} \hspace{0.5mm}
                            \includegraphics[width=8.1cm]{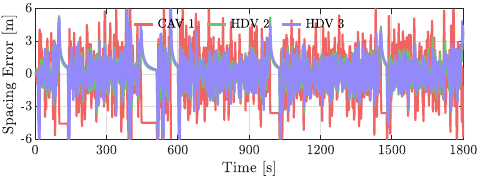}}\\
    \vspace{-1.3mm}
    \subfigure[RNDDPC]{\includegraphics[width=8.1cm]{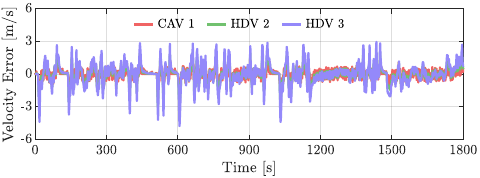}  \hspace{0.5mm}
                            \includegraphics[width=8.1cm]{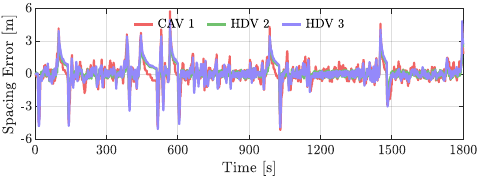}}\\
    \caption{Tracking velocity and spacing errors in comprehensive simulations for RNDDPC and baseline methods. The red line, the green line, and the purple line represent the CAV (labeled 1), the HDV (labeled 2), and the HDV (labeled 3), respectively. All data-driven control methods use the same dataset.}
	\label{Fig:ResultTracking}
	\vspace{-2mm}
\end{figure*}

The simulation results under WLTP are presented in~\figurename~\ref{Fig:ResultTracking}. For clarity, we only visualize the results of all-HDVs, linear MPC, DeePC, and proposed RNDDPC. Here, we focus on analyzing the tracking velocity and inter-vehicle spacing errors within a mixed vehicle platoon consisting of one CAV and two HDVs. Overall, in the presence of adverse factors, linear MPC in~\figurename~\ref{Fig:ResultTracking}(b) and RNDDPC in~\figurename~\ref{Fig:ResultTracking}(d) demonstrate reduced velocity and spacing errors compared to the all-HDVs scenario. This improvement is attributed to the optimal control frameworks employed by these methods, which enhance tracking performance. In contrast, the DeePC method exhibits the poorest tracking accuracy, as shown in~\figurename~\ref{Fig:ResultTracking}(c). This poor performance is likely due to its reliance on historical trajectory data to construct an online dynamic model. Such reliance makes DeePC vulnerable to noise, disturbance, and attacks, which corrupt the data and undermine its predictive accuracy and control effectiveness.

In addition, a closer examination of~\figurename~\ref{Fig:ResultTracking}(b) and (d) reveals that the red, green, and purple lines corresponding to RNDDPC show lower tracking errors compared to linear MPC. In particular, the tracking error profiles for the CAV, represented by the red line, stand out prominently. This indicates that the proposed RNDDPC outperformed linear MPC in reducing tracking errors for both CAVs and HDVs. The superior performance of RNDDPC is attributed to their ability to capture the nonlinear dynamics of the mixed vehicle platoon system, whereas linear MPC employs a simplified linear predictor. Additionally, RNDDPC utilizes a set-based approach that anticipates noise, disturbance, and attack effects, making it a more robust control strategy. Consequently, our RNDDPC ensures more stable and efficient operation of the mixed vehicle platoon system.

To further quantify the simulation results for our method and all baseline methods, we employ three key performance indices: the tracking velocity mean error $R_\mathrm{v}$, tracking spacing mean error $R_\mathrm{s}$, and total real cost $R_\mathrm{c}$ indices, as defined in~\eqref{Eq:VMADvalue}, \eqref{Eq:SMADvalue}, and~\eqref{Eq:COSTvalue}. These metrics allow a comprehensive assessment of the control effectiveness across different methods. As illustrated in Table~\ref{tab:experiment_A}, all approaches, except for the DeePC method, outperform the all-HDVs scenario in terms of tracking and control efficiency. Notably, the nonlinear MPC and Koopman MPC methods, which account for the system's nonlinear characteristics, outperform the linear MPC method. This advantage likely stems from their enhanced predictive capabilities regarding the system's future states. Regarding robustness, the ZPC method performs better than other baseline methods when subjected to adverse factors. It is noteworthy that the proposed RNDDPC method achieves the lowest values across $R_\mathrm{v}$, $R_\mathrm{s}$, and $R_\mathrm{c}$, owing to its careful consideration of nonlinear dynamics and robustness in mixed vehicle platoon control. Specifically, compared to the best-performing ZPC method among the baseline approaches, the proposed RNDDPC method still realizes further reductions of $3.6\%$, $3.3\%$, and $3.7\%$ in indices $R_\mathrm{v}$, $R_\mathrm{s}$, and $R_\mathrm{c}$, respectively. These results demonstrate that by employing the Koopman operator to capture the nonlinear characteristics of the mixed vehicle platoon and utilizing the reachable set analysis to proactively address adverse factors, our RNDDPC method can significantly improve the tracking control performance of mixed vehicle platoons with nonlinear dynamics in environments where noise, disturbances, and attacks coexist.

\begin{table*}[!ht]
\centering

\caption{Performance comparison in comprehensive simulations under non–time-delay attacks with a platoon size of $3$}
\label{tab:experiment_A}
\begin{threeparttable}
\setlength{\tabcolsep}{10pt}
\begin{tabular}{cccccccc}
\toprule
    &All HDVs &NMPC &LMPC &KMPC &DeePC &ZPC &RNDDPC\\ \midrule
$R_\mathrm{v}$   &0.81 &0.56  &0.57 &0.56 &1.24 &0.55 &\textbf{0.53}   \\ 
$R_\mathrm{s}$ &0.95 &0.61  &0.70 &0.63 &1.65 &0.60 &\textbf{0.58} \\
$R_\mathrm{c}$ &7.12 &3.07  &3.84 &3.12 &15.49 &2.97 &\textbf{2.86} \\ \bottomrule
\end{tabular}
\end{threeparttable}
\vspace{-1mm}
\end{table*}

To evaluate computational complexity, the per-step average computational time $R_\mathrm{t}$ of each algorithm is recorded across all simulations. The all-HDV benchmark, which requires no optimization, achieves the lowest $R_\mathrm{t}$ of $\SI{0.01}{ms}$. At the other extreme, nonlinear MPC exhibits the highest computational burden, averaging $\SI{38}{ms}$ due to the need to solve a nonconvex optimization problem. In comparison, both linear MPC and Koopman MPC, formulated as standard quadratic programming (QP) programs, require only $\SI{1}{ms}$ and $\SI{7}{ms}$ per step, respectively. DeePC, although also reducible to a QP, involved an expanded decision-variable space that increases the computational time to $\SI{24}{ms}$. ZPC requires $\SI{18}{ms}$ per step, as it performs online reachable-set computation. Similarly, the proposed RNDDPC method also involves online reachable-set computation, resulting in an average computational time of $\SI{22}{ms}$. Importantly, this value remains well below the simulation step size of $\SI{50}{ms}$, ensuring that RNDDPC introduces no control delays and is fully feasible for real-time implementation in mixed vehicle platoons.

\subsection{Emergency Simulations}
\label{Sec:5.3}	
\begin{figure*}[!ht]
	\centering
	\subfigure[All-HDVs]{\includegraphics[width=8.1cm]{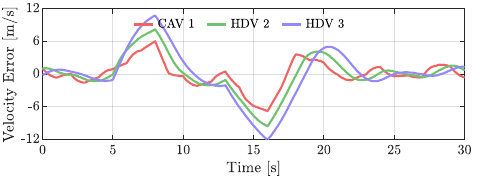} \hspace{0.5mm}
                         \includegraphics[width=8.1cm]{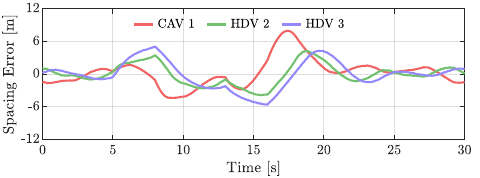}}\\
    \vspace{-1.3mm}
	\subfigure[Linear MPC]{\includegraphics[width=8.1cm]{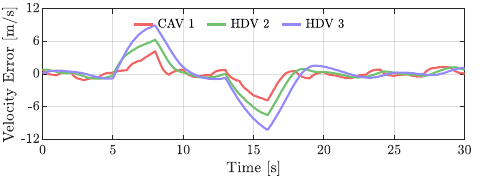} \hspace{0.5mm}
                           \includegraphics[width=8.1cm]{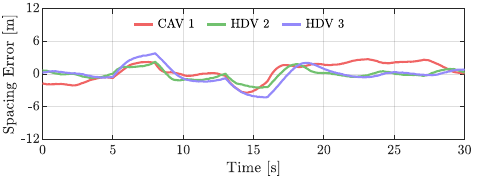}}\\
    \vspace{-1.3mm}
    \subfigure[DeePC]{\includegraphics[width=8.1cm]{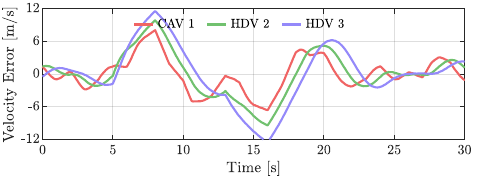} \hspace{0.5mm}
                            \includegraphics[width=8.1cm]{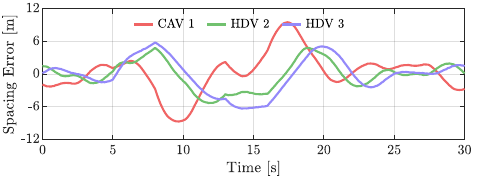}}\\
    \vspace{-1.3mm}
    \subfigure[RNDDPC]{\includegraphics[width=8.1cm]{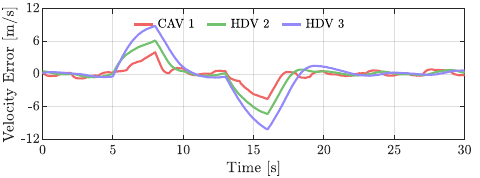}  \hspace{0.5mm}
                            \includegraphics[width=8.1cm]{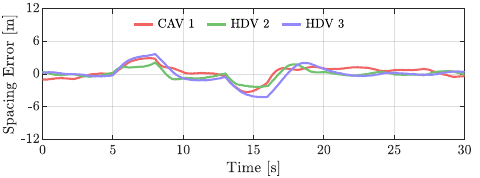}}\\
    \vspace{-2mm}
    \caption{Tracking velocity and spacing errors in emergency simulations for RNDDPC and baseline methods. The red line, the green line, and the purple line represent the CAV (labeled 1), the HDV (labeled 2), and the HDV (labeled 3), respectively. All data-driven control methods use the same dataset.}
	\label{Fig:ResultBraking}
	\vspace{-2mm}
\end{figure*}

To thoroughly assess the tracking safety of the proposed RNDDPC method, we conducted a comparative experiment under emergency conditions, as shown in~\figurename~\ref{Fig:WLTP}(b). In this scenario, the head vehicle initially travels at a constant velocity of $ \SI{20}{m/s} $, then applies an abrupt braking maneuver with a deceleration of $ \SI{-5}{m/s^2} $ for $ \SI{3}{s} $, reducing its velocity to $ \SI{5}{m/s} $. The head vehicle maintains this reduced velocity for $ \SI{5}{s} $ before accelerating at a maximum rate of  $ \SI{5}{m/s^2} $ to return to the original velocity of $ \SI{20}{m/s} $. The total duration of this emergency scenario is $\SI{30}{s}$. Simulation results for this emergency event, using different control strategies, are presented in~\figurename~\ref{Fig:ResultBraking}.

In assessing the performance of various control strategies, we employ the same evaluation metrics as in the comprehensive simulations, specifically tracking velocity error and tracking spacing error, to compare the effectiveness of each method. The results reveal that, in the presence of adverse factors, the all-HDV scheme (see~\figurename~\ref{Fig:ResultBraking}(a)) and the DeePC method (see~\figurename~\ref{Fig:ResultBraking}(c)) exhibit significant tracking errors in both velocity and spacing, indicating a lack of robustness for both approaches. While the linear MPC performs adequately in controlling velocity error, it exhibits considerable deficiencies in spacing error, particularly between $ \SI{20}{s} $ to $ \SI{30}{s} $ in~\figurename~\ref{Fig:ResultBraking}(b), underscoring the limitations of linear MPC in controlling mixed vehicle platoons. In contrast, RNDDPC demonstrates superior performance in both tracking velocity and spacing, as illustrated in~\figurename~\ref{Fig:ResultBraking}(d). These findings are consistent with the results observed in the comprehensive simulation scenarios.

Additionally, we assess the three key performance metrics, and the results for each method under emergency conditions are presented in Table~\ref{tab:experiment_B}. The analysis reveals that the proposed RNDDPC method outperforms all baseline approaches, yielding the lowest values for $R_\mathrm{v}$, $R_\mathrm{s}$, and $R_\mathrm{c}$. These results highlight the superior robustness and control precision of RNDDPC in emergency scenarios. Furthermore, they underscore the effectiveness of our approach, which leverages the Koopman operator to model the nonlinear dynamics of the mixed vehicle platoon, while proactively mitigating adverse factors such as noise, disturbances, and attacks through reachability analysis. This confirms the safety performance of our method.

\begin{table*}[ht]
\centering
\caption{Performance comparison in emergency simulations under non–time-delay attacks with a platoon size of $3$}
\label{tab:experiment_B}
\begin{threeparttable}
\setlength{\tabcolsep}{10pt}
\begin{tabular}{cccccccc}
\toprule
    &All HDVs &NMPC &LMPC &KMPC &DeePC &ZPC &RNDDPC\\ \midrule
$R_\mathrm{v}$    &2.47 &1.54  &1.58 &1.56 &3.09 &1.50 &\textbf{1.47}   \\ 
$R_\mathrm{s}$    &1.75 &0.97  &1.17 &1.01 &2.38 &0.96 &\textbf{0.94} \\ 
$R_\mathrm{c}$    &33.22 &15.25  &16.19 &15.35 &51.82 &14.49 &\textbf{14.12} \\ \bottomrule
\end{tabular}
\end{threeparttable}
\vspace{-1mm}
\end{table*}

\subsection{Simulations under Time-Delay Attacks}
In this part, we further investigate the robustness of the proposed RNDDPC algorithm by evaluating its performance under time-delay attacks. All experimental settings are identical to those in the simulations with non–time-delay attacks presented in~Section~\ref{Sec:5.2} and~Section~\ref{Sec:5.3}, except that the non–time-delay attacks are replaced by time-delay attacks. The time-delay attack is modeled as $\vartheta(k) = u(k-\tau) - u(k)$ in~\eqref{Eq:DynamicsModelCAV_L} and~\eqref{Eq:DynamicsSystem}, where the random time-delay $\tau \in \{1,2,3,4,5,6\}$ steps, corresponding to an upper bound of $\SI{0.3}{s}$. The results for comprehensive and emergency scenarios are summarized in Table~\ref{tab:experiment_C} and Table~\ref{tab:experiment_D}, respectively.

\begin{table*}[ht]
\centering

\caption{Performance comparison in comprehensive simulations under time-delay attacks with a platoon size of $3$}
\label{tab:experiment_C}
\begin{threeparttable}
\setlength{\tabcolsep}{10pt}
\begin{tabular}{cccccccc}
\toprule
    &All HDVs &NMPC &LMPC &KMPC &DeePC &ZPC &RNDDPC\\ \midrule
$R_\mathrm{v}$  &0.54 &0.47  &0.50 &0.48 &0.58 &0.47 &\textbf{0.46}   \\ 
$R_\mathrm{s}$  &0.53 &0.48  &0.52 &0.49 &0.56 &0.47 &\textbf{0.45} \\
$R_\mathrm{c}$  &3.11 &2.20  &2.39 &2.26 &3.49 &2.12 &\textbf{2.08} \\ \bottomrule
\end{tabular}
\end{threeparttable}
\vspace{-1mm}
\end{table*}

\begin{table*}[ht]
\centering
\caption{Performance comparison in emergency simulations under time-delay attacks with a platoon size of $3$}
\label{tab:experiment_D}
\begin{threeparttable}
\setlength{\tabcolsep}{10pt}
\begin{tabular}{cccccccc}
\toprule
    &All HDVs &NMPC &LMPC &KMPC &DeePC &ZPC &RNDDPC\\ \midrule
$R_\mathrm{v}$    &1.98 &1.43  &1.46 &1.45 &2.16 &1.41 &\textbf{1.38}   \\ 
$R_\mathrm{s}$    &1.23 &0.78  &0.81 &0.79 &1.47 &0.77 &\textbf{0.76} \\ 
$R_\mathrm{c}$    &24.11 &14.18  &14.98 &14.62 &28.04 &13.62 &\textbf{12.95} \\ \bottomrule
\end{tabular}
\end{threeparttable}
\vspace{-1mm}
\end{table*}

As shown in Table~\ref{tab:experiment_C} and Table~\ref{tab:experiment_D}, RNDDPC consistently outperforms baseline methods. In the comprehensive scenario, RNDDPC achieves relative reductions of $8.0\%$, $13.5\%$, and $13.0\%$ in $R_\mathrm{v}$, $R_\mathrm{s}$, and $R_\mathrm{c}$ compared with the classical linear MPC. In the more challenging emergency scenario, RNDDPC still demonstrates superior robustness, with reductions of $5.5\%$, $6.2\%$, and $13.6\%$ on the same indices. Although ZPC achieves competitive results in certain cases, it remains inferior to RNDDPC, while DeePC performs the worst under time-delay attacks. These findings confirm that RNDDPC maintains strong control performance even in the presence of time-delay attacks. It is worth emphasizing that, although RNDDPC is originally designed for non-time-delay attacks, its explicit treatment of system uncertainty through reachability analysis allows it to effectively manage the predictive uncertainty induced by time-delay attacks, thereby ensuring robust performance across diverse adversarial conditions.

\subsection{Simulations under Different Platoon Sizes}
To investigate the scalability of the proposed framework, we develop a flexible configuration mechanism for the Koopman network to accommodate mixed vehicle platoons of different sizes. As shown in~\figurename~\ref{Fig:KoopmanDNN}, the input and output dimensions of each sub-network in the Koopman network depend solely on the platoon size $n$ and the lifted state dimension $n_\mathrm{z}$. A key step in achieving scalability is establishing a clear and systematic relationship between $n$ and $n_\mathrm{z}$. In this study, we define the lifted state dimension as a linear function of the platoon size, i.e., $n_\mathrm{z} = 4n$, enabling the Koopman network to adjust seamlessly to various platoon sizes without requiring a complete architectural redesign.

We evaluate this design through simulation studies conducted under emergency scenarios for different platoon configurations. Following the recommendation in~\cite{zhou2021analytical} that platoons typically consist of up to six vehicles in mixed traffic, we extend the simulation with $n=3$ vehicles in Section~\ref{Sec:5.3} to include cases with $n=4$ and $n=5$ vehicles. The corresponding results are summarized in Table~\ref{tab:experiment_E} and Table~\ref{tab:experiment_F}. A comparison with Table~\ref{tab:experiment_B} shows that all control methods exhibit increasing values of the performance indices $R_\mathrm{v}$, $R_\mathrm{s}$, and $R_\mathrm{c}$ as the platoon size increases. This trend reflects the reduced influence of the leading CAV on vehicles at the tail of the platoon, making precise control more challenging. Despite this inherent difficulty, the proposed RNDDPC consistently demonstrates superior performance across all platoon sizes. These results highlight the robustness of RNDDPC and the effectiveness of the Koopman network configuration mechanism. Furthermore, these results confirm that RNDDPC can be systematically scaled to larger mixed vehicle platoons without significant architectural modifications, underscoring its strong potential for deployment in real-world traffic systems.

\begin{table*}[!ht]
\centering
\caption{Performance comparison in emergency simulations under non-time-delay attacks with a platoon size of $4$}
\label{tab:experiment_E}
\begin{threeparttable}
\setlength{\tabcolsep}{10pt}
\begin{tabular}{cccccccc}
\toprule
    &All HDVs &NMPC &LMPC &KMPC &DeePC &ZPC &RNDDPC\\ \midrule
$R_\mathrm{v}$    &2.58 &1.73  &1.77 &1.74 &3.17 &1.71 &\textbf{1.69}   \\ 
$R_\mathrm{s}$    &1.79 &1.04  &1.19 &1.06 &2.44 &1.03 &\textbf{1.01} \\ 
$R_\mathrm{c}$    &34.93 &19.01  &19.97 &19.17 &53.83 &18.40 &\textbf{18.01} \\ \bottomrule
\end{tabular}
\end{threeparttable}
\vspace{-1mm}
\end{table*}

\begin{table*}[!ht]
\centering
\caption{Performance comparison in emergency simulations under non-time-delay attacks with a platoon size of $5$}
\label{tab:experiment_F}
\begin{threeparttable}
\setlength{\tabcolsep}{10pt}
\begin{tabular}{cccccccc}
\toprule
    &All HDVs &NMPC &LMPC &KMPC &DeePC &ZPC &RNDDPC\\ \midrule
$R_\mathrm{v}$    &2.97 &2.12  &2.16 &2.13 &3.37 &2.08 &\textbf{2.05}   \\ 
$R_\mathrm{s}$    &1.86 &1.20  &1.29 &1.22 &2.46 &1.18 &\textbf{1.16} \\ 
$R_\mathrm{c}$    &38.87 &24.55  &25.47 &24.59 &54.81 &23.79 &\textbf{22.87} \\ \bottomrule
\end{tabular}
\end{threeparttable}
\vspace{-1mm}
\end{table*}

\section{Conclusions}
\label{Sec:6}
This paper presents a novel Robust Nonlinear Data-Driven Predictive Control (RNDDPC) method for mixed vehicle platoons. This method explicitly addresses the nonlinear dynamics inherent in mixed vehicle platoons and is designed to ensure robustness against adverse factors, including noise, disturbances, and attacks. To handle the system's nonlinearity, we employ a deep EDMD learning technique that maps the system's nonlinear behavior into a linear representation in a high-dimensional space, providing a more tractable model for control design. To further enhance model robustness, we introduce an over-approximated matrix zonotope set method, which enables secondary learning to construct a Koopman-based reachable set predictor, effectively compensating for potential adverse factors. By integrating Koopman operator theory with reachability analysis technology, we formulate the RNDDPC optimization problem to achieve robust control, even in the presence of adverse factors, thus guaranteeing system safety. Compared to existing methods, extensive simulation results demonstrate the superior performance and robustness of the proposed method in mixed vehicle platoon control.

Our work has some limitations that require further research. First, the current framework does not consider communication delays, assuming that the cloud control platform can access real-time data from the mixed vehicle platoon. Extending the proposed approach to account for communication delays, represents an important direction for future work. Second, the present study focuses solely on longitudinal car-following scenarios and does not address the dynamic behaviors associated with HDVs joining or exiting the platoon. Addressing the challenges introduced by switching communication topologies in such scenarios would further enhance the scalability and practical applicability of the proposed control framework. Finally, the training data and algorithm validation are conducted entirely within a high-fidelity simulation environment. Future work should include field experiments to construct real-world datasets and carry out full-scale vehicle testing.

\section*{Acknowledgement}
The work of S.~Li, Q.~Xu, J.~Wang, and K.~Li is supported by the National Natural Science Foundation of China with 52221005, and the National Natural Science Foundation of China with 52072212. K. Yang is supported by the Singapore Ministry of Education (MOE) under its Academic Research Fund Tier 1 (A-8003262-00-00).

\appendix
\section{Proof of Lemma 1}
\label{Appendix:Appendix A}
To simplify the derivations, the sequences of unknown modeling errors are defined as:
	\begin{equation}
	O_-=[\sigma(1),\sigma(2),\ldots,\sigma(T)] \in \mathbb{R} ^{n_\mathrm{z} \times T},
	\end{equation}    
	\begin{equation}
	\Gamma_-=[\varrho(1),\varrho(2),\ldots,\varrho(T)] \in \mathbb{R} ^{2n \times T},
	\end{equation}    
although it is important to note that $O_-$ and $\Gamma_-$ are not measurable.

For the equivalent Koopman-based model in~\eqref{Eq:KoopmanDynamicsRobust}, the following relationships hold:
\begin{equation}
Z_{+} = \begin{bmatrix}A~B~H~J\end{bmatrix}\begin{bmatrix}
Z_{-} \\
U_{-} \\
E_{-} \\
F_{-} 
\end{bmatrix} + O_{-},
\end{equation}
\begin{equation}
X_{+} = C Z_{+} + \Gamma_{-}.
\end{equation}
Given that the matrices $ \begin{bmatrix}Z_{-}^{\top}~U_{-}^{\top}~E_{-}^{\top}~F_{-}^{\top}\end{bmatrix}^{\top} $ and $ \begin{bmatrix}
Z_{+}
\end{bmatrix} $ are of full row rank, we can derive the following expressions:
\begin{equation}
\begin{bmatrix}A~B~H~J\end{bmatrix} =\left(Z_{+}-O_{-}\right)\begin{bmatrix}
Z_{-} \\
U_{-} \\
E_{-} \\
F_{-} 
\end{bmatrix}^{\dagger},
\end{equation}
\begin{equation}
C =\left(X_{+}-\Gamma_{-}\right)\begin{bmatrix}
Z_{+}
\end{bmatrix}^{\dagger},
\end{equation}
where the sequences $O_{-}$ and $\Gamma_{-}$ are unknown, but by applying the corresponding matrix zonotope sets $\mathcal{M}_{\sigma}$ and $\mathcal{M}_{\varrho}$ in~\eqref{Eq:setofnoise_1} and~\eqref{Eq:setofnoise_2}, we can obtain the matrix zonotope sets $\mathcal{M}_{\mathrm{ABHJ}}$ and $\mathcal{M}_{\mathrm{C}}$ in~\eqref{Eq:M_ABHJ} and~\eqref{Eq:M_C}. These matrix zonotope sets provide an over-approximation of the system models that effectively accounts for the modeling errors.

\section{Proof of Lemma 2}
\label{Appendix:Appendix B}
For the equivalent Koopman-based model in~\eqref{Eq:KoopmanDynamicsRobust}, the state reachable set can be calculated using the following model:    
\begin{equation}
\begin{cases}
\mathcal{R}^\mathrm{z}{(i+1|k)}=[A~B~H~J]\left(\mathcal{R}^\mathrm{z}{(i|k)} \times \mathcal{Z}_{u} \times \mathcal{Z}_{\epsilon} \times \mathcal{Z}_{\vartheta} \right) +\mathcal{Z}_{\sigma}, \\
{\mathcal{R}}{(i+1|k)}=C \mathcal{R}^\mathrm{z}{(i+1|k)} +\mathcal{Z}_{\varrho},
\end{cases}
\end{equation}
where $\begin{bmatrix}A~B~H~J\end{bmatrix} \in \mathcal{M}_{\mathrm{ABHJ}}$ and $\begin{bmatrix}C\end{bmatrix} \in \mathcal{M}_{\mathrm{C}}$, as stated in Lemma~\ref{Lemma:M_ABHJ}. Assuming that $ \mathcal{R}^\mathrm{z}{(i|k)} $ and $\hat{\mathcal{R}}^\mathrm{z}{(i|k)}$ originate from the same initial set, it follows that ${\mathcal{R}}{(i+1|k)} \subset \hat{\mathcal{R}}{(i+1|k)}$. Consequently, the recursive relation in~\eqref{Eq:ReachableSet} holds, providing an over-approximated reachable set for the system state.

\section{Recursive Feasibility}
\label{Appendix:Appendix C}
In this part, we demonstrate that the proposed RNDDPC optimization problem in~\eqref{Eq:RNDDPCOptimizationProblem} is recursively feasible. This result is established based on~Assumption~\ref{Assumption_Terminal}, Definition~\ref{Definition:PropagationSet}, Lemma~\ref{Lemma:Monotonicity}, and Lemma~\ref{Lemma:Shifted}, and is formally stated in Lemma~\ref{Lemma:Feasibility}.

\begin{assumption}[Terminal Backup Feasibility]
\label{Assumption_Terminal}
The constraint set $\mathcal{X}$ is one-step robustly control invariant under some admissible backup input $u_\mathrm b\in\mathcal{U}$.
\end{assumption}

\begin{definition}[Set Propagation Operators]
\label{Definition:PropagationSet}
For a set $S \subset \mathbb{R}^{n_\mathrm{z}}$ and an input $u\in\mathcal{U}$, define
\begin{equation}
\mathcal{T}(S,u) \triangleq \mathcal{M}_{\mathrm{ABHJ}}\big(S \times \langle u,0\rangle \times \mathcal{Z}_\epsilon \times \mathcal{Z}_\vartheta\big)+\mathcal{Z}_\sigma,\qquad
\mathcal{O}(S)\triangleq \mathcal{M}_{\mathrm{C}}S+\mathcal{Z}_\varrho.
\end{equation}
Using these operators, constraints~\eqref{Eq:48b}--\eqref{Eq:48c} can be compactly rewritten as
\begin{equation}
\hat{\mathcal{R}}^{\mathrm z}(i+1|k)=\mathcal{T}\!\left(\hat{\mathcal{R}}^{\mathrm z}(i|k),u(i|k)\right),\qquad
\hat{\mathcal{R}}(i+1|k)=\mathcal{O}\!\left(\hat{\mathcal{R}}^{\mathrm z}(i+1|k)\right).
\end{equation}
\end{definition}

\begin{lemma}[Monotonicity of Propagation]
\label{Lemma:Monotonicity}
For any sets $S_1\subseteq S_2$ and input $u\in\mathcal{U}$,
\begin{equation}
\mathcal{T}(S_1,u)\subseteq \mathcal{T}(S_2,u), \qquad
\mathcal{O}(S_1)\subseteq \mathcal{O}(S_2).
\end{equation}
\end{lemma}
\textbf{Proof:} The operators $\mathcal{M}_{\mathrm{ABHJ}}(\cdot)$ and $\mathcal{M}_{\mathrm{C}}(\cdot)$ are affine and monotone with respect to set inclusion. Cartesian products with fixed compact sets and Minkowski sums with fixed zonotopes preserve inclusion. Therefore, the stated inclusions hold.

\begin{lemma}[Shifted Candidate and Nested Reachable Sets]
\label{Lemma:Shifted}
Assume that the RNDDPC optimization problem~\eqref{Eq:RNDDPCOptimizationProblem} is feasible at time $k$, yielding feasible input sequences $\{u^\star(i|k)\}_{i=0}^{N-1}$ and sets $\{\hat{\mathcal{R}}^{\mathrm z}(i+1|k),\hat{\mathcal{R}}(i+1|k)\}_{i=0}^{N-1}$. Apply $u(k)=u^\star(0|k)$ to the system, and denote the measured state at the next time step as $x(k+1)$. Then
\begin{equation}
\hat{\mathcal{R}}(0|k+1)=\langle x(k+1),0\rangle \subseteq \hat{\mathcal{R}}(1|k).
\end{equation}
Next, consider the shifted input sequence at time $k+1$:
\begin{equation}
\tilde u(i|k+1)=
\begin{cases}
u^\star(i+1|k), & i=0,\dots,N{-}2,\\
u_\mathrm{end}, & i=N-1,
\end{cases}
\end{equation}
where $u_\mathrm{end}\in\mathcal{U}$ is an admissible terminal control input as required by Assumption~\ref{Assumption_Terminal}. Then the corresponding propagated sets satisfy, for $i=0,\dots,N-1$,
\begin{equation}
\hat{\mathcal{R}}^{\mathrm z}(i|k+1)\ \subseteq\ \hat{\mathcal{R}}^{\mathrm z}(i+1|k),
\qquad
\hat{\mathcal{R}}(i|k+1)\ \subseteq\ \hat{\mathcal{R}}(i+1|k).
\end{equation}
\end{lemma}
\textbf{Proof:} Feasibility at time $k$ implies $x(k+1)\in \hat{\mathcal{R}}(1|k)$ by construction of~\eqref{Eq:48b}--\eqref{Eq:48e}. Hence $\hat{\mathcal{R}}(0|k+1)=\langle x(k+1),0\rangle\subseteq \hat{\mathcal{R}}(1|k)$. Using the same inputs $u^\star(1|k),\dots,u^\star(N-1|k)$ for the first $N-1$ steps and the monotonicity in Lemma~\ref{Lemma:Monotonicity}, we obtain
\begin{equation}
\hat{\mathcal{R}}^{\mathrm z}(1|k+1)=\mathcal{T}\big(\hat{\mathcal{R}}^{\mathrm z}(0|k+1),u^\star(1|k)\big)
\subseteq \mathcal{T}\big(\hat{\mathcal{R}}^{\mathrm z}(1|k),u^\star(1|k)\big)=\hat{\mathcal{R}}^{\mathrm z}(2|k),
\end{equation}
and inductively $\hat{\mathcal{R}}^{\mathrm z}(i|k+1) \subseteq \hat{\mathcal{R}}^{\mathrm z}(i+1|k)$ for $i=1,\dots,N-1$. Applying $\mathcal{O}(\cdot)$ and Lemma~\ref{Lemma:Monotonicity} again yields $\hat{\mathcal{R}}(i|k+1) \subseteq \hat{\mathcal{R}}(i+1|k)$ for $i=1,\dots,N-1$.

\begin{lemma}[Recursive Feasibility of RNDDPC]
\label{Lemma:Feasibility}
If the RNDDPC optimization problem~\eqref{Eq:RNDDPCOptimizationProblem} is feasible at time $k$, it remains feasible at time $k+1$.
\end{lemma}
\textbf{Proof:} Construct the shifted input sequence $(\tilde u(i|k+1))_{i=0}^{N-1}$ as in Lemma~\ref{Lemma:Shifted}. For $i=0,\dots,N-2$, $\tilde u(i|k+1)=u^\star(i+1|k)\in\mathcal{U}$ by feasibility at time $k$. For the terminal control input, choose $u_\mathrm{end}=u_\mathrm b$ so that one-step propagation from $\hat{\mathcal{R}}^{\mathrm z}(N-1|k+1)$ keeps $\hat{\mathcal{R}}(N|k+1)= \mathcal{O}\!\big(\mathcal{T}(\hat{\mathcal{R}}^{\mathrm z}(N-1|k+1),u_{\mathrm b})\big) \subseteq \mathcal{X}$. Furthermore, based on Lemma~\ref{Lemma:Shifted}, for all $i=0,\dots,N-1$ we have $\hat{\mathcal{R}}(i|k+1) \subseteq \hat{\mathcal{R}}(i+1|k)\subseteq \mathcal{X}$, hence~\eqref{Eq:48d} holds at time $k+1$. By construction, $\hat{\mathcal{R}}(0|k{+}1)=\langle x(k+1),0\rangle$ satisfies~\eqref{Eq:48h}, and $x(i+1|k+1) \in \hat{\mathcal{R}}(i+1|k+1)$ can be enforced exactly as in~\eqref{Eq:48e}. Input constraints~\eqref{Eq:48f} hold by the choice of $\tilde u$ and $u_\mathrm{end}\in\mathcal{U}$. Therefore all constraints~\eqref{Eq:48b}–\eqref{Eq:48h} are satisfied at time $k+1$, i.e., the problem is feasible.

\section{Closed-Loop Constraint Satisfaction}
\label{Appendix:Appendix D}

In this part, we present the lemma and its proof that establish closed-loop constraint satisfaction within the proposed RNDDPC optimization problem in~\eqref{Eq:RNDDPCOptimizationProblem}.

\begin{lemma}[Closed-Loop Constraint Satisfaction]        
	\label{Lemma:Stability}
	Consider the system described in~\eqref{Eq:KoopmanDynamicsRobust}, subject to the input and output constraints~\eqref{Eq:46c} and~\eqref{Eq:46d}. Let the disturbances $\epsilon(k)$, attacks $\vartheta(k)$, and modeling errors $\sigma(k)$ and $\varrho(k)$ in~\eqref{Eq:KoopmanDynamicsRobust} be bounded by the zonotope sets $\mathcal{Z}_{\epsilon}$, $\mathcal{Z}_{\vartheta}$, $\mathcal{Z}_{\sigma}$, and $\mathcal{Z}_{\varrho}$ in~\eqref{Eq:ReachableSet}.
    If the RNDDPC problem~\eqref{Eq:RNDDPCOptimizationProblem} is feasible at time steps $k$, then the resulting data-driven controller ensures that the closed-loop system~\eqref{Eq:KoopmanDynamicsRobust} robustly satisfies the constraints in~\eqref{Eq:46c} and~\eqref{Eq:46d} at all time steps $k$, thereby guaranteeing closed-loop constraint satisfaction.
\end{lemma}
\textbf{Proof:} According to Lemma~\ref{Lemma:ReachableSet} and its proof in~\ref{Appendix:Appendix B}, the computed reachable sets $\hat{\mathcal{R}}{(i|k)}$ constitute over-approximations of the reachable sets ${\mathcal{R}}{(i|k)}$. At each time step $k$, the control input $u^{*}(i|k)$ is chosen such that all constraints~\eqref{Eq:48b}–\eqref{Eq:48h} are satisfied. This ensures that the system state $x^{*}(i|k)$ remains within the intersection of the over-approximated reachable set $\hat{\mathcal{R}}{(i|k)}$ and the state constraint set $\mathcal{X}$, and $u^{*}(i|k)$ satisfies the input constraint $\mathcal{U}$. Consequently, under the feasibility of the RNDDPC problem in Lemma~\ref{Lemma:Feasibility}, the data-driven controller derived from~\eqref{Eq:RNDDPCOptimizationProblem} guarantees that the closed-loop system~\eqref{Eq:KoopmanDynamicsRobust} robustly satisfies both state and input constraints at every time step $k$. This confirms the closed-loop constraint satisfaction of the system under the proposed control framework.

\section{Extraction of Uncertainty Features}
\label{Appendix:Appendix E}
Due to the presence of noise in the data, modeling errors are unavoidable. Based on the equivalent Koopman-based model in~\eqref{Eq:KoopmanDynamicsRobust}, the modeling errors $\sigma(k)$ and $\varrho(k)$ can be expressed as:
\begin{equation}
\label{Eq:KoopmanDynamicsErrors}
\begin{cases}
\sigma(k)=  z(k+1) - A z(k)-B {u}(k)-H \epsilon(k)- J\vartheta(k), \\
\varrho(k) = {x}(k+1) - C z(k+1),
\end{cases}
\end{equation}
where $ A $, $ B $, $ H $, $ J $, $ C$, and the lifted state $z(k)$ are learned from the Koopman-based model and are therefore known.

\begin{figure*}[ht]
	\centering
\subfigure[{Probability distribution of $\sigma(k)$}]{\includegraphics[width=8cm]{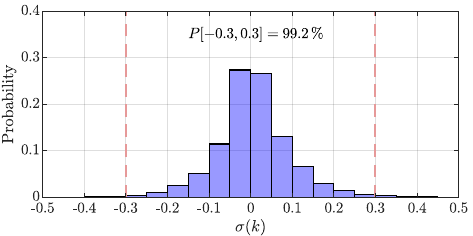}}
\subfigure[{Probability distribution of $\varrho(k)$}]{\includegraphics[width=8cm]{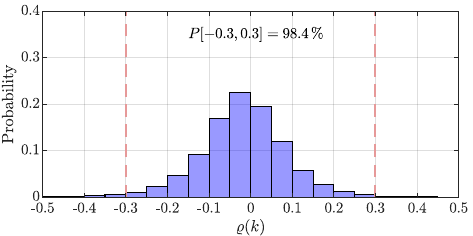}}\\
\caption{Probability distribution of modeling errors $\sigma(k)$ and $\varrho(k)$.}
	\vspace{-0.1cm}
\label{Fig:KoopmanErrors}
\end{figure*}

Using~\eqref{Eq:KoopmanDynamicsErrors} and collected data sequences in~\eqref{Eq:DataSequences}, the probability distributions of each element value in $\sigma(k)$ and $\varrho(k)$ are obtained, as illustrated in~\figurename~\ref{Fig:KoopmanErrors}. These distributions characterize the statistical properties of the modeling errors and provide a quantitative basis for selecting appropriate error bounds in~\eqref{Eq:W_Bound}. To balance robustness and conservatism in controller design, we set $\sigma_\mathrm{max}=\mathbf{1}_{n_\mathrm{z}} \otimes 0.3$ and $\varrho_\mathrm{max}=\mathbf{1}_{2n} \otimes 0.3$ in the simulations, which include $99.2\%$ and $98.4\%$ of the uncertainties, respectively.

\section{Quantifying the Over-Approximation of Matrix Zonotope Sets}
\label{Appendix:Appendix F}
In this part, we quantitatively evaluate the tightness of the over-approximation introduced by matrix zonotope sets. As defined in~\eqref{Eq:M_ABHJ} and~\eqref{Eq:M_C}, the matrix zonotope sets $\mathcal{M}_{\mathrm{ABHJ}}$ and $\mathcal{M}_{\mathrm{C}}$ are constructed to approximate the uncertainty of the learned Koopman system matrices $\begin{bmatrix}A~B~H~J\end{bmatrix}$ and $\begin{bmatrix}C\end{bmatrix}$. Specifically, these sets $\mathcal{M}_{\mathrm{ABHJ}}$ and $\mathcal{M}_{\mathrm{C}}$ capture the admissible variation ranges of each element in the corresponding system matrices. The width of such ranges directly reflects the degree of approximation tightness. To quantify this property, we compute the deviation between the upper and lower bounds of the matrix zonotope sets and the nominal values of the Koopman matrices. The deviation measures are defined as:
\begin{equation}
\label{Eq:MatrixZonotopeError_ABHJ}
\begin{cases}
e_{\mathrm{ABHJ}}^{\mathrm{sup}} =  \mathcal{M}_{\mathrm{ABHJ}}^{\mathrm{sup}} - [A~B~H~J], \\
e_{\mathrm{ABHJ}}^{\mathrm{inf}} =  \mathcal{M}_{\mathrm{ABHJ}}^{\mathrm{inf}} - [A~B~H~J],
\end{cases}
\end{equation}
\begin{equation}
\label{Eq:MatrixZonotopeError_C}
\begin{cases}
e_{\mathrm{C}}^{\mathrm{sup}} =  \mathcal{M}_{\mathrm{C}}^{\mathrm{sup}} - [C], \\
e_{\mathrm{C}}^{\mathrm{inf}} =  \mathcal{M}_{\mathrm{C}}^{\mathrm{inf}} - [C],
\end{cases}
\end{equation}
where $\mathcal{M}_{\mathrm{ABHJ}}^{\mathrm{sup}}$ and $\mathcal{M}_{\mathrm{ABHJ}}^{\mathrm{inf}}$ denote the element-wise upper and lower bound matrices of $\mathcal{M}_{\mathrm{ABHJ}}$, respectively. Similarly, $\mathcal{M}_{\mathrm{C}}^{\mathrm{sup}}$ and $\mathcal{M}_{\mathrm{C}}^{\mathrm{inf}}$ represent the element-wise bound matrices of $\mathcal{M}_{\mathrm{C}}$. The resulting deviation matrices $e_{\mathrm{ABHJ}}^{\mathrm{sup}}$ and $e_{\mathrm{ABHJ}}^{\mathrm{inf}}$ describe the upper and lower over-approximation deviations of $\mathcal{M}_{\mathrm{ABHJ}}$ with respect to $\begin{bmatrix}A~B~H~J\end{bmatrix}$, while $e_{\mathrm{C}}^{\mathrm{sup}}$ and $e_{\mathrm{C}}^{\mathrm{inf}}$ characterize the corresponding deviations of $\mathcal{M}_{\mathrm{C}}$ with respect to $\begin{bmatrix}C\end{bmatrix}$.

\begin{figure*}[ht]
	\centering
\subfigure[Probability distribution of all elements in $e_{\mathrm{ABHJ}}^{\mathrm{sup}}$ and $e_{\mathrm{C}}^{\mathrm{sup}}$]{\includegraphics[width=8.1cm]{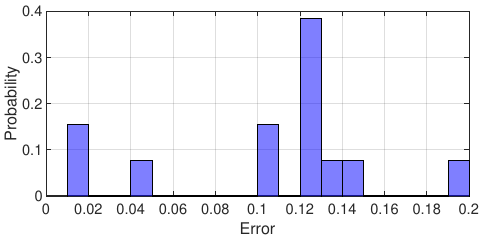}}
\subfigure[Probability distribution of all elements in $e_{\mathrm{ABHJ}}^{\mathrm{inf}}$ and $e_{\mathrm{C}}^{\mathrm{inf}}$]{\includegraphics[width=8.1cm]{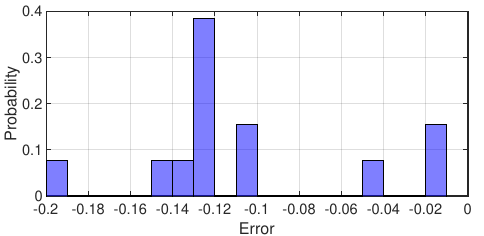}}\\
\caption{{Probability distribution of the over-approximation deviations.}}
	\vspace{-0.1cm}
\label{Fig:MatrixZonotopeError}
\end{figure*}

By applying~\eqref{Eq:MatrixZonotopeError_ABHJ} and~\eqref{Eq:MatrixZonotopeError_C}, the deviation matrices $e_{\mathrm{ABHJ}}^{\mathrm{sup}}$, $e_{\mathrm{ABHJ}}^{\mathrm{inf}}$, $e_{\mathrm{C}}^{\mathrm{sup}}$, and $e_{\mathrm{C}}^{\mathrm{inf}}$ are obtained, which provide a rigorous quantification of the approximation accuracy. The statistical distributions of all elements in $e_{\mathrm{ABHJ}}^{\mathrm{sup}}$ and $e_{\mathrm{C}}^{\mathrm{sup}}$ are presented in~\figurename~\ref{Fig:MatrixZonotopeError}(a), while those for $e_{\mathrm{ABHJ}}^{\mathrm{inf}}$ and $e_{\mathrm{C}}^{\mathrm{inf}}$ are shown in~\figurename~\ref{Fig:MatrixZonotopeError}(b). As expected, all upper over-approximation deviations are strictly greater than zero, whereas all lower over-approximation deviations are strictly less than zero. This observation confirms that the matrix zonotope sets $\mathcal{M}_{\mathrm{ABHJ}}$ and $\mathcal{M}_{\mathrm{C}}$ indeed serve as valid over-approximations of the learned Koopman system matrices $\begin{bmatrix}A~B~H~J\end{bmatrix}$ and $\begin{bmatrix}C\end{bmatrix}$. Furthermore, from the perspective of approximation tightness, the maximum upper deviation reaches $0.2$, and the maximum lower deviation is $-0.2$, indicating that the over-approximation bounds are not excessively conservative.

\bibliography{Reference}

\vfill
\end{document}